\documentclass[a4paper,fleqn,usenatbib]{mnras}

\usepackage{newtxtext}
\usepackage{newtxmath}
\usepackage{adjustbox,lipsum}
\usepackage{caption}
\usepackage{multirow}

\usepackage[T1]{fontenc}
\usepackage{ae,aecompl}
\usepackage{ae,aecompl}
\usepackage[figuresright]{rotating}
\usepackage{float}
\usepackage{array}

\usepackage{threeparttable}
\usepackage{lscape}
\usepackage{upgreek}
\usepackage{graphicx}	
\usepackage{amsmath}	
\usepackage{amssymb}	
\usepackage{comment}
\usepackage{lineno}
\usepackage{color}
\usepackage[toc,page]{appendix}
\usepackage{verbatim}
\usepackage{booktabs}
\usepackage{multirow}
\usepackage{siunitx}

\newcommand{\ttwo}{\textsc{tempo2}}
\newcommand{\tn}{\textsc{temponest}}
\newcommand{\mn}{\textsc{Multinest}}

\newcommand{\trs}{\mathbfss{T}}
\newcommand{\cov}{\mathbfss{C}}
\newcommand{\desm}{\mathbfss{D}}
\newcommand{\FI}{\mathbfss{J}}
\newcommand{\tres}{\boldsymbol{t}}

\newcommand{\bsy}{\boldsymbol}
\newcommand{\delm}{\delta{}m}
\newcommand{\diag}{\textrm{diag}}
\newcommand{\dr}{IPTA DR 1}
\newcommand{\chm}{CHM10}

\newcommand{\glc}{GLC18}

\newcommand{\Ms}{M_{\odot}}

\newcommand{\GMsRatioIPTANom}{(GM)_{\textrm{IPTA1}}^{\textrm{SSE}}/(\mathcal{GM})_{\odot}^{\textrm{N}}}
\newcommand{\GMsRatioIPTAumoNom}{(Gm)_{\textrm{IPTA1}}^{\textrm{SSE}}/(\mathcal{GM})_{\odot}^{\textrm{N}}}
\newcommand{\GMsRatioIAUNom}{(GM)_{\textrm{IAU}}^{\textrm{CBE}}/(\mathcal{GM})_{\odot}^{\textrm{N}}}
\newcommand{\CHMIPTAratio}{\sigma_{\textrm{\small{CHM10}}}/\sigma_{\textrm{\small{IPTA1}}}}
\newcommand{\IPTAIAUratio}{\sigma_{\textrm{\small{IPTA1}}}/\sigma_{\textrm{\small{IAU}}}}
\usepackage{pifont}
\usepackage{upgreek}
\title[Short title, max. 45 characters]{MNRAS \LaTeXe\ template -- title goes here}

\title[Solar System studies with the IPTA]{Studying the solar system with the International Pulsar Timing Array}
\author[R. N. Caballero et al.]{\parbox{\textwidth}{{\large R.~N.~Caballero,$^{1}$\thanks{E-mail:caballero.astro@gmail.com}
Y.~J.~Guo,$^{2}$
K.~J.~Lee,$^{2,1}$\thanks{E-mail:kjlee@pku.edu.cn}
P.~Lazarus,$^{1}$
D.~J.~Champion,$^{1}$
G.~Desvignes,$^{1}$
M.~ Kramer,$^{1,3}$
K.~Plant,$^{4}$
Z.~Arzoumanian,$^{5}$
M.~Bailes,$^{6}$
C.~G.~Bassa,$^{7}$
N.~D.~R.~Bhat,$^{8}$
A.~Brazier,$^{9,10}$
M.~Burgay,$^{11}$
\mbox{S.~Burke-Spolaor,$^{12,13}$}
S.~J.~Chamberlin,$^{14}$
S.~Chatterjee,$^{10,15}$
I.~Cognard,$^{16,17}$
J.~M.~Cordes,$^{10,15}$
S.~Dai,$^{18}$
P.~Demorest,$^{19}$
T.~Dolch,$^{20}$
R.~D.~Ferdman,$^{21}$
E.~Fonseca,$^{22}$
J.~R.~Gair,$^{23}$
\mbox{N.~Garver-Daniels,$^{12,13}$}
P.~Gentile,$^{12,13}$
M.~E.~Gonzalez,$^{24,25}$
E.~Graikou,$^{1}$
L.~Guillemot,$^{16,17}$
G.~Hobbs,$^{18}$
G.~H.~Janssen,$^{7,26}$
R.~Karuppusamy,$^{1}$
M.~J.~Keith,$^{3}$
M.~Kerr,$^{27}$
M.~T.~Lam,$^{12,13}$
P.~D.~Lasky,$^{28}$
T.~J.~W.~Lazio,$^{29}$
L.~Levin,$^{12,3}$
K.~Liu,$^{1}$
A.~N.~Lommen,$^{30}$
D.~R.~Lorimer,$^{12,13}$
R.~S.~Lynch,$^{31}$
D.~R.~Madison,$^{31}$
R.~N.~Manchester,$^{18}$
J.~W.~McKee,$^{3,1}$
M.~A.~McLaughlin,$^{12,13}$
S.~T.~McWilliams,$^{12,13}$
C.~M.~F.~Mingarelli,$^{1,32}$
D.~J.~Nice,$^{33}$
S.~Os{\l}owski,$^{34,1,6}$
N.~T.~Palliyaguru,$^{35}$
T.~T.~Pennucci,$^{36}$
B.~B.~P.~Perera,$^{3}$
D.~Perrodin,$^{11}$
A.~Possenti,$^{11}$
S.~M.~Ransom,$^{31}$
D.~J.~Reardon,$^{28,18}$
S.~A.~Sanidas,$^{37,3}$
A.~Sesana,$^{38}$
G.~Shaifullah,$^{7}$
R.~M.~Shannon,$^{39,6}$
X.~Siemens,$^{40}$
J.~Simon,$^{29}$
R.~Spiewak,$^{40,6}$
I.~Stairs,$^{25}$
B.~Stappers,$^{3}$
D.~R.~Stinebring,$^{41}$
K.~Stovall,$^{19}$
J.~K.~Swiggum,$^{40}$
S.~R.~Taylor,$^{42}$
G.~Theureau,$^{16,17,43}$
C.~Tiburzi,$^{1,34}$
L.~Toomey,$^{18}$
R.~van~Haasteren,$^{29}$
W.~van~Straten,$^{44}$
J.~P.~W.~Verbiest,$^{34,1}$
J.~B.~Wang,$^{45}$
X.~J.~Zhu,$^{28,39}$
and 
W.~W.~Zhu$^{46,1}$
 }}
\vspace{0.4cm} \\ 
{\it Affiliations are listed at the end of the paper}
}

\pubyear{2018}

\begin{document}
\label{firstpage}
\pagerange{\pageref{firstpage}--\pageref{lastpage}}
\maketitle

\begin{abstract}
Pulsar-timing analyses are sensitive to errors in the solar-system ephemerides (SSEs) 
that timing models utilise to estimate the location of the solar-system barycentre, the quasi-inertial 
reference frame to which all recorded pulse times-of-arrival are referred.
Any error in the SSE will affect all pulsars, therefore pulsar timing arrays (PTAs) are
a suitable tool to search for such errors and 
impose independent constraints on relevant physical parameters.
We employ the first data release of the International Pulsar Timing Array
to constrain the masses of the planet-moons 
systems and to search for possible unmodelled 
objects (UMOs) in the solar system. 
We employ ten SSEs from two
independent research groups, derive and compare mass constraints of planetary systems,
and derive the first PTA mass constraints on asteroid-belt objects.
Constraints on planetary-system masses have
been improved by factors of up to 20 from the previous relevant study using the same assumptions,
with the mass of the Jovian system measured at 9.5479189(3)$\times10^{-4}$\,$M_{\odot}$.
The mass of the dwarf planet Ceres is measured at 4.7(4)$\times10^{-10}$\,$M_{\odot}$.
We also present the first sensitivity curves using real data that place generic limits 
on the masses of UMOs, which can also be used as upper limits on the mass
of putative exotic objects. For example, upper limits on dark-matter clumps
are comparable to published limits using independent methods.
While the constraints on planetary masses derived with all employed SSEs are consistent, 
we note and discuss differences in the associated timing residuals and UMO sensitivity curves. 
\end{abstract}




\section{Introduction}
\label{sec:intro}
Millisecond pulsars (MSPs) are the most stable rotators 
known to date in the observable Universe. 
Pulsar timing \citep[see e.g.][]{lk2005} is a powerful technique through which we record the 
times-of-arrival (TOAs) of the pulses and use a sophisticated model to convert the
topocentric TOA, or site arrival time, 
to the pulse time-of-emission in the pulsar's co-moving 
reference frame.
The success of the model's fit is assessed from the timing residuals---the
difference between the observed and the model-predicted TOAs---which
capture all the information unaccounted for in the timing model.
Timing residuals from contemporary high-precision timing of the brightest
and most stable MSPs observed 
are at levels of a few hundreds of nanoseconds \citep[see e.g.][]{vlh+2016}.

The high precision with which MSPs can be timed has made them 
the primary targets for studies of gravity in the
(quasi-stationary) strong field regime,
primarily  through the studies of their
orbital behaviour and especially when in tight
orbits with other neutron stars \citep[see e.g.][for a review]{Dam2009}.
Additionally, MSPs can
be used as reference clocks to study
interesting  phenomena
that affect their TOAs but are
extrinsic to their rotational and orbital behaviour.
It is self-evident that it is to our advantage to
use multiple MSPs to observe such extrinsic 
phenomena when possible, especially when trying
to measure an effect which is expected to affect TOAs
from all MSPs and depends on the pulsar's sky position. 
We refer to such an ensemble of regularly observed MSPs 
as a pulsar timing array \citep[PTA;][]{fb1990}. 
The primary scientific goal of PTA researchers
is the direct detection of low-frequency gravitational waves (GWs), at nHz frequencies,
including stochastic GW backgrounds (GWBs).
Three collaborations are currently leading these efforts,
namely the European Pulsar Timing Array \citep[EPTA;][]{dcl+2016},
the North-American Nanohertz Observatory for Gravitational Waves \citep[NANOGrav;][]{abb+2015a},
and the Parkes Pulsar Timing Array  \citep[PPTA;][]{rhc+2016}. 
These collaborations work together under the International Pulsar Timing Array consortium \citep[IPTA;][]{vlh+2016}
in an effort to combine data, resources and expertise to maximise their scientific output.

While the timing model includes
the pulsar's rotational, astrometric, and orbital parameters, 
and accounts for the time-delay effects of the interstellar medium on
the pulsed-signal propagation, it is in fact the transformation of
the observation site arrival time to the arrival time 
at the solar-system barycentre (SSB) that may
introduce correlated signals in the TOAs 
most likely to interfere with the GWB searches.
Such correlated signals may arise from 
errors in the terrestrial time standards and the solar-system ephemeris (SSE)
used to predict the position of the SSB at any given time of interest.
The correlated signals from these two types of errors result
in monopolar and dipolar spatial correlations, respectively \citep[see][]{thk+2016},
leading to cross-correlations in 
the timing residuals of pulsar pairs that may resemble 
those from a GWB, which have their basis on the 
quadrupole angular correlation pattern caused by GWs \citep{hd1983}. 
The presence of signals from clock and SSE errors 
increases the false-alarm 
probability of GW detection with PTAs \citep{thk+2016}.
In principle, these signals are distinguishable from each other
if the data is sufficiently informative, 
and to manage this, it is especially important
to increase the number of MSPs contributing to the analysis
and to the sampling of the cross-correlation curve \citep{sej+2013,tve+2016}.
While we examine methods to minimise
these errors and mitigate their effects when searching for GWs in the PTA data,
one can also use the data to extract scientific information on topics other than GWs. 
In particular, PTA data have been employed to develop an independent
pulsar-based time standard \citep{hcm+2012} and to constrain the masses of the 
solar-system planetary systems (SSPS) \citep[][henceforth CHM10]{chm+2010}. 

The SSEs that we use for pulsar timing are
constructed via numerical integrations 
of the equations of motion for the known
solar-system bodies. These integrations
are subject to a wealth of 
observational data from telescopes, radio and laser ranging 
and spacecrafts orbiting the planets and their moons, when available.
Such input data also include estimates of the masses of 
the planets and other important solar-system bodies.
Observationally, it is not the mass, $M$, but the gravitational parameter
of the bodies that is determined, i.e. $GM$ where $G$ is the universal constant of gravitation.
This parameter can be determined with much higher precision than the 
gravitational constant \citep[see e.g.][]{pl2010}, a fact that
limits the precision of measurements of  $M$ in SI units.
For this reason, the masses
solar-system bodies such as the planets 
are expressed as the ratios of their gravitational parameters
to the solar gravitational parameter 
(heliocentric gravitational constant), $GM_{\odot}$.

New data are added over time, so that newer
SSEs are subject to data of better accuracy and observational sampling.
Many of the involved parameters are fitted
and adjusted while creating the final SSE.
As noted in \chm{}, while this process gives accurate
predictions for the positions of the planetary-systems with
respect to the Earth-Moon system,
they do not manage to constrain the masses much better than
the measurements used as initial values.
This is reflected by the fact that typically the ratios
of the gravitational parameters of the
planetary systems with respect to the solar parameter
are held fixed during numerical integrations.
What changes between SSE versions 
with respect to the reference planetary masses
is either the initial mass values of the planetary systems, 
for example after new mass estimates by spacecraft fly-bys,
and/or the estimate of the 
solar gravitational parameter, which can be a fitted parameter in the SSE.
Therefore, the input planetary masses
in principle differ between the various SSE versions.
With this in mind, \chm{} search for errors in the 
masses of the planetary systems, as the most possible errors 
that pulsar-timing data could identify.

In this paper we focus
on extending the work of \chm{}
using the first IPTA data release \citep[\dr{};][]{vlh+2016}.
In addition to improving the constraints 
on the SSPS masses, we provide the first PTA constraints on 
the most massive asteroid-belt objects (ABO) and
employ a recently-published 
algorithm \citep*[][henceforth \glc{}]{glc2018} 
to search for possible unmodelled objects (UMOs) in Keplerian orbits 
in the solar system.
We also make a quantitative comparison of 
SSEs provided by two independent groups,
namely the Institut de M\'ecanique C\'eleste et de Calcul des \'Eph\'em\'erides (IMCCE)
and the Jet Propulsion Laboratory (JPL).

The rest of the paper is organized as follows.
In Section~\ref{sec:Dataset} we briefly overview the \dr{}
and list the MSPs used in the present study and their basic observational properties.
In Section~\ref{sec:Methods} we describe the analysis methods, which 
includes the details of single-pulsar noise analysis and analyses
for constraining the mass of the eight planetary systems and ABOs,
as well as the masses of UMOs.
The results of the analyses are presented in Section~\ref{sec:results}.
We finally discuss scientific implications of our results and our conclusions in Section~\ref{sec:Disc}.

\section[]{Data set: The \dr{}}
\label{sec:Dataset}
\dr{} is described in \cite{vlh+2016}, and we only give a 
brief overview in this section. The full data set consists of TOAs from 49 MSPs.
Data were collected by the three regional PTAs over a total time-span of 
up to 27.1~yr using seven telescopes across the world, namely
the Effelsberg Radio Telescope, the Lovell Telescope, the Nan\c cay Radio Telescope, 
and the Westerbork Synthesis Radio Telescope by the EPTA, 
the Arecibo Observatory and the Green Bank Telescope by NANOGrav,
and the Parkes Radio Telescope by the PPTA. 
The \dr{} data set was constructed by combining data that were published in 
\cite*{ktr1994}, \cite{dfg+2013}, \cite{mhb+2013}, \cite{zsd+2015}, and \cite{dcl+2016}.

It is important to note that the TOAs from the different telescopes and 
different studies were calculated with various methods. 
Although all TOA calculations were based 
on template-matching methods \citep{tay1992},
where each observed profile is cross correlated with a profile template of arbitrary phase,
there are technical differences regarding issues such as 
the methods to create the pulse-profile templates,
and algorithms for optimal template matching. 
There are also different approaches with regards to the
way that the recorded information is used. 
For example, in some cases the total intensity profiles were used,
which are created by summing the flux of all polarization modes,
frequency bands and sub-integrations, while in others cases, one TOA
was calculated per frequency band. These choices reflect 
differences in the sensitivity of instruments over time and analysis
methods which have developed to address them. 
For example, data from a receiver with limited total bandwidth
would use total intensity profiles to achieve useful signal-to-noise ratio,
while a more modern broadband receiver can achieve sufficient signal-to-noise ratio
with sub-bands of the total bandwidth. In this case, one may opt to produce
TOAs per sub-band as a way to mitigate, for example,
effects of possible evolution of the pulse profile
over the observing frequency \citep[see e.g.][]{xkj+1996,kll+1999}, 
or possible noise that is limited in certain sub-bands \citep*{lsc+2016,css2016}.
One may also opt to not integrate profiles over time for short-period
pulsars in order to better sample the orbit \citep[e.g.][]{dcl+2016}.
 
For MSPs for which data from more than one PTA 
were available, the IPTA data combination 
increased the time-span of the MSP data, as well as their cadence
and observing-frequency coverage. 
Increased time-span and cadence
allows improved sampling of orbits
at longer and shorter periods, respectively. 
They also lead to better characterisation of low- and high-frequency
noise properties. Noise mitigation
is further aided by 
improved observing-frequency coverage which is particularly crucial 
in measuring chromatic noise processes related with the turbulent 
ionised interstellar medium. 
The combination of data from multiple telescopes, when available, 
also offers the chance to use individual data sets in the same
observing-frequency bands to search for noise due to systematics \citep{lsc+2016}.

The \dr{} served as a first testing ground for
the use of pulsar-timing noise models that were more
complex compared to previous 
studies such as  
\cite{abb+2015a}, \cite{cll+2016}, or \cite{rhc+2016},
which only used data from individual PTAs.
It was exactly the aforementioned properties of the \dr{}
that motivated the inclusion of additional noise components in the 
noise analysis presented in \cite{lsc+2016}. The analysis
was made in particular to attempt to distinguish
between noise specific to each pulsar
and noise due to systematics in the data of a given observing system,
or noise that is associated with a specific observing frequency band.
The intent of introducing the latter noise term is
to probe chromatic noise that does not
follow the dispersive law of cold homogeneous plasma, associated
with temporal dispersion measure (DM) variations \citep[see e.g.][]{kcs+2013,lbj+2014}.

In the present paper we study the timing data from six MSPs in total
and employed data from five of these 
to constrain the masses of solar-system bodies.
The MSPs were selected based on 
the contribution of each MSP to the overall results as discussed in Section~\ref{sec:psrSelect}.
The key observational properties of the data for each of these
pulsars are presented in Table~\ref{tab:MSPs}. 
By comparison to the \dr{} data, the one change we have made is related 
to PSR~J1713+0747. The large number of TOAs (19972) would
make the current analysis significantly computationally expensive.
This large number of TOAs primarily stems from NANOGrav data,
which are not averaged 
over the observing frequency band, resulting in one TOA per frequency channel.
To reduce the computational cost for PSR~J1713+0747
we employed the \ttwo{} routine \textsc{AverageData} and produced an 
average TOA for each epoch per observing frequency band by summing up 
all channels across the frequency band. 
\begin{table}
\small
\caption{General characteristics of the \dr{}.0 data \citep{vlh+2016}
for the MSPs used in this study 
(note that PSR~J0437$-$4715 was not used to derive mass limits of solar-system bodies; see Section~\ref{sec:psrSelect}). 
For each pulsar we note the total time-span, $T$, the average cadence, 
the number of telescopes contributing data,
and the weighted root-mean-square (RMS) of the timing residuals
(after subtracting the waveform of the DM variations)
The residual RMS was derived using the planetary ephemeris DE421.
}
\begin{minipage}{0.2\textwidth}
\tabcolsep=0.11cm
\begin{tabular}{l c c c c}
\toprule
PSR       & $T$           & Average   &  Number of  & Residual \\
Name     &              & cadence         &  telescopes & weighted RMS        \\
(J2000)  &   (yr)    & (d)                  &                    &  ($\mu$s)  \\
\midrule
J0437$-$4715 & 14.9 & 5.1  & 1 & 0.3\\
J0613$-$0200 & 13.7 & 4.3  & 6 & 1.2\\
J1012+5307 & 14.4 & 6.3  & 5 & 1.7\\
J1713+0747 & 17.7 & 5.1  & 7 & 0.3\\
J1744$-$1134 & 17.0 & 8.4  & 6 & 1.1\\
J1909$-$3744 & 10.8 & 4.4 & 3 & 0.2\\
\bottomrule
\end{tabular}
\label{tab:MSPs}
\end{minipage}%
\end{table}

\section[]{Analysis methods}
\label{sec:Methods}

We have implemented two methods to study the solar system with the \dr{}.
Both methods rely on searching for residuals induced by the periodic oscillation
of the SSB due to the presence of a mass in orbit that is not accounted for by the
pulsar timing models. This mass can either be a difference from the real mass
of a solar-system body to that assumed by the SSE,
for which we employed the method discussed in Section~\ref{sec:DJC10}, 
or the mass of a UMO not included in the SSE, 
for which we employed the method discussed in Section~\ref{sec:bayes}. 

We clarify here, that in this study 
we are only modelling possible errors in the SSE reference masses.
We do not examine the effects of positional errors.
Under this model, possible small errors in orbital elements
could be absorbed in the mass-error parameter.
As we noted on Section~\ref{sec:intro}, 
we expect that mass errors 
are more likely to be detected first with pulsar timing analysis,
however, sensitivity to errors in orbital elements are not
excluded. The \glc{} method can also be focused on applying
upper limits on orbital parameter of UMOs, but this is beyond
of the scope of this study.
We discuss further work in pulsar-timing research
that attempts to extend PTA studies to orbital elements
of planets in Section~\ref{sec:Disc}.

Prior to discussing the 
SSE analysis, we first
give an overview of the single-pulsar timing and noise analysis.

\subsection[]{Single-pulsar timing and noise analysis}
\label{sec:SPNA}

As with other applications of PTAs,
constraining the masses of known or unknown
bodies in orbit around the SSB requires 
good characterisation of the noise 
in individual pulsar data \citep[see][for reviews on sources of noise in pulsar timing]{cor2013,vs2018}, 
as noise components may have significant power
at frequencies related to a planetary orbit.
Insufficient accounting of the noise can lead to
significant bias on the measured values of the timing parameters
and their uncertainties \citep{chc+2011,vl2013}.
\chm{} pointed out these effects in the 
context of constraining planets' masses and specifically
did not include one of the four pulsars they used, 
PSR~J0437$-$4715, when
estimating the mass error of Mars. 
Specifically, \chm{} argued 
that its noise model was not sufficient to
account for spectral features close to the orbital
frequency of Mars, and including this pulsar
would thus bias the solution for the specific planet.

For the work presented in this paper,
for each pulsar we created 
different timing and noise models 
for each SSE. The initial 
phase-coherent timing models
were obtained using
the timing software \ttwo{} \citep*{hem2006}.
\ttwo{} uses a previously derived timing model (which could be as simple
as the pulsar discovery position and rotational frequency)
and iteratively performs a least-squares fit of the model to
the TOAs until the reduced chi-squared of the residuals is minimized. 
\ttwo{} applies a linearised approximation 
to calculate the small, linear offsets of model parameters
from the pre-fit value \citep*[see also][]{ehm2006}.
The least-squares fit can be unweighted or weighted according to the TOA uncertainties. 
Throughout this work, our timing solutions use weighted fits.
These initial individual pulsar-timing models
do not include parameters related to 
errors in the SSE or any noise components.
We then employed \tn{} \citep{lah+2014} to perform a Bayesian
(simultaneous) timing and noise analysis, 
with the same noise modelling used, for example, in \cite{cll+2016}.
\tn{} samples the joint parameter space
of the timing and noise parameters using \mn{} \citep{fhb2009}, 
a Bayesian inference algorithm based on nested sampling \citep[see][]{ski2004},
while evaluating the timing model at each point of the parameter space using the \ttwo{} algorithms.

Before proceeding to the correlated-signal analysis,
we produced the final noise models employing the analysis package 
that we use to make the search for errors in the SSE in order to 
have a consistent mathematical noise-model parametrization.
During this last stage, 
we performed a Bayesian noise analysis while
analytically marginalising over the timing parameters, also using \mn{} as the sampler.
In brief, the noise model consists of the following components:
\begin{itemize}
\item{
Uncorrelated noise terms, modelled with a pair of 
corrections to the TOA uncertainties per observing system (white noise). 
The \tn{} analysis includes an EFAC (for Error FACtor, a multiplicative factor)
and an EQUAD (for Error in QUADrature, a factor added in quadrature).
The application of these terms attempts to create a timing solution where 
appropriate relative weights between the different observing systems are given,
since TOA uncertainties calculated via template-matching methods,
do not always fully account for the TOA scatter.
EFACs are used to correct underestimation of the uncertainty, for example due to low
signal-to-noise ratio of the observed pulse profile, differences in the pulse profile and the template or
presence of noise other than white radiometer noise in the profile, at significant levels. EQUADs are
primarily used to account for additional scatter in the TOAs due to physical processes such 
as pulse phase jitter \citep[e.g.][]{lkl+2012,sod+2014}. 
The corrected TOA uncertainty, $\hat{\sigma}$, and initial uncertainty, $\sigma$, 
are then related as
\begin{equation}
\label{eq:sigma}
\hat{\sigma}^2 =  (\sigma\cdot{}\textrm{EFAC})^2+\textrm{EQUAD}^2
\end{equation}
During the final noise analysis, we applied a `global' EFAC per pulsar, to regularize the white-noise level
against the other noise components.
}
\item{
An achromatic (observing-frequency independent)  
low-frequency stochastic
component (red noise) per pulsar,
modelled as a wide-sense stationary 
stochastic process with a one-sided 
power-law spectrum of the form
\begin{equation}
\label{eq:pspec}
S(f) = \frac{A^2}{f} \left(\frac{f}{f_\textrm{c}} \right)^{2\tilde{\alpha}}\,, 
\end{equation}
where $f$ is the Fourier frequency,
 $f_\textrm{c}=1\, \textrm{yr}^{-1}$ is a reference frequency,
$A$ is the amplitude in units of time,
and $\tilde{\alpha}$ is the spectral index.
This noise component is added to model primarily physical noise from
irregularities in the rotation of the pulsar, often referred to as `spin noise' \citep[e.g.][]{sc2010,klo+2006}.
In the absence of other dedicated model components \citep[see][]{lsc+2016}, this component
will also include noise due to possible systematics in the data. 
}
\item{
A chromatic (observing-frequency dependent) 
low-frequency stochastic component (DM noise). 
It has the same spectral properties as the red-noise component,
but with the restriction that the induced residuals 
reflect TOA delays that follow
the dispersive law of cold homogeneous plasma \citep[e.g.][]{ll1960}, 
i.e. the time delay of a signal
at two observing frequencies, $\nu_1$ and $\nu_2$, 
along a line-of-sight
with DM value, $D_{l}$, is
\begin{equation}
\label{eq:dmlaw}
\Delta{}T_{D_{l}} = \kappa\frac{D_{l}}{\textrm{pc}\,\textrm{cm}^{-3}}\left[\left(\frac{\nu_{1}}{\textrm{GHz}}\right)^{-2} - \left(\frac{\nu_{2}}{\textrm{GHz}}\right)^{-2} \right]\,,
\end{equation}
where $\kappa = 4.15\times10^{-3}$~s.
}
\end{itemize}

The power-law power spectra used to describe
the stochastic noise components have sharp
cut-offs at $f=1/T$, with $T$ the data span. 
This cut-off reflects the fact that power
at frequencies below $1/T$ is fitted out by the timing
model, as discussed in previous works \citep{vlm+2009,lbj+2014}.
In particular, we fit for the rotational period and period derivative
to remove the low-frequency power of the red noise,
and the DM first and second derivatives to remove 
the low-frequency power of the DM-variations noise.
As such, a linear and a quadratic term 
for DM-variations are always implemented in the
(deterministic) timing models used in this work.

Finally, we note that the timing model also needs to take into account
the dispersive delays from the plasma of the solar wind \citep{yhc+2007b}.
Our timing models implement the standard \ttwo{} solar-wind model \citep{ehm2006},
that assumes a spherical distribution of free electrons
with a nominal density of 4\,cm$^{-3}$ at 1\,AU.
Deviations of the electron density distribution from this value 
(e.g. due to solar activity or deviations from the assumed electron density distribution 
and/or density at 1\,AU) 
will induce additional delay signals that become significant when the line-of-sight
to the pulsar is close the to the solar disc. 
The result is then induced residuals with annual signatures, 
which peak at epochs where the pulsar is at its smallest elongation.
Pulsars with low ecliptic latitudes are more susceptible to such effects
(only PSR~J1744$-$1134 falls into that category from the pulsars used in this study).
Unmitigated solar-wind signals have complex power spectra 
and show spatial correlations similar to that caused by SSE errors 
\citep{thk+2016} so that they could interfere with the sensitivity
of PTA data at high frequencies. More careful modelling of the 
solar wind is planned for future work. Data from new
observing campaigns at lower frequencies \citep[see e.g.][]{tv2018} can provide
valuable input for better modelling and mitigation of 
dispersive delays from the solar wind.

\subsection[]{Analysis method for known solar-system bodies}
\label{sec:DJC10} 

We first discuss our approach in
searching for coherent waveforms in the MSPs
from possible errors in the SSPS masses assumed in the SSE. 
We employ a frequentist analysis using a code 
that implements the method described in \chm{}.

The method considers small errors, $\delm{}$, in SSPS masses, $m$,
so that $\delm{} \ll m$. Such errors will induce residuals
due to periodic linear shifts in the SSB position with the period of the
planetary orbit. 
In such a small mass-error case, we can neglect 
higher-order effects on the residuals due to the SSB motion.
\chm{} examined the extent of secondary effects
using a modified version of the DE421 SSE
where the mass of Jupiter deviated the real value by 7$\times10^{-11}\Ms$ 
(an amount compatible to the precision that current PTAs can probe the Jovian system mass, as one can
see from the results in the next Section) 
and concluded that such effects were negligible 
in the case of Jupiter after fitting for the timing model.
We further investigated these secondary effects with methods
similar to the work in \chm{} and reached similar conclusions.
The cases of the inner planets, Mercury and Venus, show 
additional complexity because of the effects on the orbit of the Earth-Moon system
that errors in these planetary masses would cause. The induced residuals
from such effects, however, fall into different frequencies to the 
orbital frequencies of the inner planets. 
Consequently, although a fully dynamical model
could make use of such signals as additional information
in constraining the planetary masses,
we have verified that these signals, if present in the data, 
do not affect the results and conclusions from
the narrow-frequency signal search
employed in this work.

In the first-order \chm{} approximation, the induced residuals from the erroneous mass
are then only associated 
with the (solar-system related) R\o{}mer delay, 
the geometric vacuum delay of the TOA at the observatory and at the SSB.
The induced residuals will reflect the
shift in the position of the SSB along the
barycentric position vector of the SSPS, $\bsy{b}$, associated with
an error in the pulse time-of-emission. 
For the multi-pulsar and multi-SSPS case,
this error is calculated for each time epoch as
\begin{equation}
\label{eq:dmass1}
\tau^{n,k}_{b}\approx \frac{1}{cM_{\rm{T}}} \sum_{i,j}^{n,k}\delm{}_{i}(\bsy{b}_{i}\cdot\hat{\bsy{R}}_{\textrm{j}})\,,
\end{equation}
where indices i and j refer to the i-th (out of $n$) SSPS and the j-th (out of $k$) pulsar, respectively,
$\bsy{b}$ is the barycentric position vector of the SSPS,
$\hat{\bsy{R}}_{\textrm{j}}$ is the unit barycentric position vector of the pulsar
(or pulsar binary) barycentre,
 $c$ is the speed of light and $M_{\rm{T}}$ is the total mass
of the solar system, 
which was approximated by $M_{\rm{T}}\approx{}\Ms{}$.

Since Eq.~\eqref{eq:dmass1} is linear, $\tau_{b}$ can be
directly added to the linear timing model of \ttwo{}.
Although a single pulsar can provide
measurements of the $\delm{}$ parameters,
Eq.~\eqref{eq:dmass1} shows how the measurement 
precision is dependent on the pulsar's sky position
and, therefore, better and less biased measurements can be
made by fitting for these parameters simultaneously
with many pulsars. 
In \chm{}, this was performed using 
\ttwo{}, which was appropriately modified 
to allow the $\delm{}$ parameters to 
be fitted in a `global' timing analysis.
Such an analysis is a simultaneous fit of
the timing models of various pulsars, where a subset of the 
parameters, which we call global, are common for all pulsars.
In this example, the $\delm$ parameters are the global parameters.

The covariance matrix for each pulsar, $\cov$, is
constructed using the maximum-likelihood values 
of the posterior distribution of the Bayesian noise analysis.
It is defined as
\begin{equation}
\label{eq:covM}
\cov = \cov{}_{\textrm{w}} + \cov{}_{\textrm{r}} + \cov{}_{\textrm{d}}\,,
\end{equation}
where the constituent matrices are the white-, red- and DM-noise 
covariance matrices. $\cov{}_{\textrm{w}}$ is a diagonal matrix with 
the main diagonal populated with the variances of the TOAs
(after application of EFACs and EQUADs). 
The red- and DM-noise
covariance matrices are populated by elements defined, respectively, as \citep{lbj+2014}
\begin{equation}
\label{eq:covMr}
 \cov{}_{\textrm{r},ij} =\int_{1/T}^\infty S_\textrm{r}(f) \cos(2\pi ft_{ij}) \textrm{d}f\,, \textrm{and}
\end{equation}
\begin{equation}
\label{eq:covMdm}
\cov{}_{\textrm{dm},ij} =\frac{\kappa^2\int_{1/T}^\infty S_\textrm{d}(f) \cos(2\pi ft_{ij}) \textrm{d}f}{\nu_i^2\nu_j^2}\,.
\end{equation}
In the above equations, the $i$ and $j$ indices refer to observing epochs,
and $\nu$ denotes the observing frequency.

We can now proceed to search for coherent waveforms
as predicted by Eq.~\eqref{eq:dmass1} via a global timing analysis. 
During our analysis, apart from the global parameters,
for each pulsar we only fitted for a limited number of
timing parameters to ensure that the condition 
numbers of the design matrices (discussed below)
were small and matrix inversions are computationally stable.
The timing parameters fitted for are the rotational frequency and its derivative, the 
DM and derivatives (first and second included in timing models), 
the pulsar position and parallax. 
The rotational frequency, DM, and their derivatives 
correlate with low-frequency noise parameters
and $\delm$ parameters related to the  planets with the longest periods.
Pulsar position and parallax are also significantly
affected by changes to the SSEs
\citep[see also Fig.~1 in][]{cab2018}.
We have done so after confirming that the
timing models were not influenced by this practice.

Using standard linear-algebra methods we fitted the timing parameters 
denoted with the column matrix, $\epsilon$, as
\begin{equation}
\label{eq:LinAlg1}
\epsilon = (A_{\textrm{r}}^{\trs}A_{\textrm{r}})^{-1}A_{\textrm{r}}^{\trs{}}A_{\textrm{q}}^{\trs}\cov{}^{-\frac{1}{2}}\tres{}\,, 
\end{equation}
and the corresponding variances are given by
\begin{equation}
\label{eq:LinAlg2}
\sigma_{\epsilon}^2 = \diag(\FI{}^{-1})\,, \
\FI{} = \desm{}^{\trs}\cov{}^{-1}\desm{}
\end{equation}
In these equations, $\tres{}$ is the column matrix of the timing residuals,
$\desm{}$ is the design matrix (calculated with \ttwo{} during  the individual 
pulsar timing analysis), $\cov{}$ is the covariance matrix 
and $\FI{}$ is the Fisher-information matrix. 
$A_{\textrm{q}}$ and $A_{\textrm{r}}$
the Q and R decompositions of matrix $A=\cov{}^{-\frac{1}{2}}\desm{}$, respectively.
The $^{\trs{}}$,$^{-1}$ and $^{-\frac{1}{2}}$ superscripts
denote the transpose, inverse and inverse of the square root of a given matrix, respectively. 
All matrices are the total matrices, for all pulsars;
$\tres{}$ is formed by appending all pulsar timing residuals,
$\cov{}$ is the block-diagonal matrix of all pulsar covariance matrices, 
and \desm{} is formed by appending the SSPS-waveform column matrices
to the block-diagonal matrix of all pulsar design matrices. In this way, the 
SSPS-waveforms act as global parameters to the fit.

The columns with the global SSPS $\delm$ waveforms are calculated
using Eq.~\eqref{eq:dmass1}. The position of the pulsar is known from the timing model.
The position vector of the SSPS for a given observing epoch is 
calculated based on the information for the SSPS orbits 
provided by the used SSE. 
IMCCE and JPL
provide libraries that contain 
modules and functions that read in the data
from the ephemerides and calculate the positions and velocities of the SSPSs
for given times. IMCCE and JPL provide the
\textsc{calceph}\footnote{http://www.imcce.fr/fr/presentation/equipes/ASD/inpop/calceph/} \citep{glm+2015}
and \textsc{spice}\footnote{https://naif.jpl.nasa.gov/naif/toolkit.html}
libraries respectively. 
Having confirmed that
both libraries give completely consistent results,
we used the \textsc{calceph} in all related work,
except the calculations regarding 
mass errors of ABOs, as discussed in Section~\ref{sec:ResKnownAst}. 

\subsection[]{Analysis method for unknown solar-system bodies}
\label{sec:bayes}

The approximation used in \chm{} can also be employed 
in the case where instead of errors in the SSE's reference 
mass of the SSPS, 
we consider the mass of UMOs, for which we then also need to
model the dynamics of their motion.
Such an analysis is beneficial for different reasons.
Firstly, it gives the potential to PTAs to probe the masses and dynamics
of any object in orbit around the SSB and to impose
constraints on physical parameters of proposed or hypothetical objects (see Section~\ref{sec:ResUnKnown}), 
such as Planet Nine \citep{bb2016} or dark matter in the solar system
with specified mass distributions \citep{lb2005,ppA2013,pAp2013}.
In this study we focus on a simple model which assumes
small bodies in Keplerian orbits around the SSB in order to 
probe to first order the sensitivity of the real PTA data set to orbiting masses in the solar system.
While not specifically applied in order to constraint the parameter space of specific proposed objects,
the analysis assumes orbits that approximate those of most solar-system bodies
(excluding perturbations) and the results can serve as 
a confirmation of our mass constraints on known bodies and as a means to compare 
the different SSEs at first order.

For this analysis, we implement the algorithm 
presented in \glc{}, which
searches for coherent waveforms from bodies in Keplerian orbits around the SSB,
in the TOAs of all pulsars.
The details of the approach to search for UMOs,
including the mathematical framework, the choice of prior distributions and
the analysis algorithm, can be found in \glc{}.
The algorithm solves the dynamical problem of bodies in Keplerian orbits.
By neglecting higher-order effects due to the SSB motion as in 
\chm{} and any perturbations on the UMO from any object except the Sun,
the algorithm is currently restricted to searches of small objects and that 
are not in orbit around a major planet.
The dynamical model contains seven unknown parameters, 
i.e. the mass of the UMO, $m$, and
the six Keplerian orbital parameters, i.e.
the semi-major axis, $a$, the eccentricity, $e$, 
the longitude of the ascending node, $\Omega$,
the inclination of the orbit, $i$, 
the argument of perihelion, $\omega$, and the reference phase, $\phi_{0}$.
For a set of values for these parameters, the model
determines the barycentric position vector of the UMO, $\bsy{b}$, 
and uses Eq.~\eqref{eq:dmass1} to calculate the induced signal in the TOAs
$S(\xi)$, where we use $\xi$ to denote the seven unknown parameters.

The UMO-induced waveform is now an unknown waveform in the data, and no longer part
of the timing parameters. The analysis now uses the reduced likelihood \citep{vlm+2009},
which is used when solving the problem while analytically marginalising over
the parameters that are not of interest (often referred to as nuisance parameters).
In this case, these are all the timing parameters, $\boldsymbol{\epsilon}$.
For the multi-pulsar case, where we search for a coherent waveform $S$ in all pulsars, 
the reduced likelihood function can be written as
\begin{equation}
\label{eq:reducedLikFunc2}
\begin{split}
\Lambda\propto&  \frac{1}{\sqrt{|\rm \cov\cov'|}} \times \\
  & \exp\left(-\frac{1}{2} \sum_{i,j,I,J} ({ 
	\tres}_{I,i}-S(\xi)_{I,i})^{\trs} {C'}_{I,J, i, j} ({\tres}_{J,j}-S(\xi)_{J,j}\right)\,,
\end{split}	
\end{equation}
where the $I,J$ indices denote pairs of pulsars, the $i,j$ indices denote pairs of time epochs,
and 
${\cov{}'}={\cov{}}^{-1}-{\cov{}}^{-1}{\desm{}}({\desm{}}^{\textrm{{\trs{}}}}{\cov{}}^{-1}{\desm{}})^{-1}{\desm{}}^{\textrm{\trs{}}}{\cov{}}^{-1}$.
We note that one can use
the alternative formulation of the likelihood introduced in \cite{lah+2013}.
By applying Bayes's theorem, one can proceed to 
perform Bayesian parameter estimation as
\begin{equation}
\label{eq:Binfer}
P(\zeta|X)\propto P(\zeta)\Lambda\,.
\end{equation}
In this compact notation, $X$ is the data
and $\zeta$ are all the
model parameter we want to sample,
that is, the Keplerian orbital parameters of the UMO
and the pulsar-noise parameters we opt to fit simultaneously.
Therefore, $P(\zeta|X)$ is the posterior probability distribution of the parameter(s) of interest,
and $P(\zeta)$ is the prior probability distribution of the parameter(s).
The parameter space is explored using \mn{}.

The analysis algorithm for UMOs offers
flexibility in the analysis, allowing analytical
marginalization over the timing parameters and
limiting the prior range of orbital parameters or fixing them to a given 
value. For the work presented in this paper,
we analytically marginalize over the timing parameters
and simultaneously search over the UMO orbital
parameters and pulsar-noise parameters.

Following the same procedure as in \glc{},
we first ran an analysis using the least informative
priors for the parameters in order to get the 
posterior distributions from which we can determine
whether we have a possible detection of a UMO.
These priors are uniform in
the log-space for the parameters with dimension
and uniform for dimensionless parameters.
In the non-detection case, as is the case in all our \dr{} analyses,
we proceeded to a follow-up, upper-limit analysis to determine the
data's sensitivity to any given UMO mass at any semi-major axis value.
For the upper-limit analysis, we changed the mass priors to uniform in linear space
and performed Bayesian inference for a grid of fixed semi-major axis values.
We will refer to these upper limits of the mass as a function of the semi-major axis
as the mass sensitivity curves. 

\section[]{Analyses and Results}
\label{sec:results}

The analysis with our implementation of the \chm{} method used ten different SSEs, 
five from IMCCE (designation ``INPOP'') and five from JPL (designation ``DE''). 
An overview of the SSEs we employed can be found in Table~\ref{tab:SSEs}.
Before proceeding to searches for correlated SSE-error signals across pulsars,
we performed some preliminary searches for errors in masses of SSPSs
using single-pulsar data to check the effects of the noise model we select 
and to compare the performance of our implementations of the \chm{} method
with that of \ttwo{}.
We also made a first-order comparison of the effects on pulsar timing
from choosing a different SSE during the analysis. 
\begin{table}
\small
\caption{List of SSEs used in the analyses.}
\begin{tabular}{lc}
\toprule
IMCCE Ephemerides & Reference\\
\midrule
INPOP06C  &\cite{fml+2008} \\
INPOP08  & \cite{flm+2009} \\
INPOP10E & \cite{fml+2013} \\
INPOP13C  &\cite{fml+2014} \\
INPOP17A & \cite{vfg+2017} \\
\midrule
JPL Ephemerides & Reference\\
\midrule
DE405 &  \cite{sta1998}\\
DE418 &  \cite{fsw+2007}\\
DE421 &  \cite{fwb2009}\\
DE430 &  \cite{fwb+2014}\\
DE435 &  \cite{fpj+2016}\\
\bottomrule
\end{tabular}
\label{tab:SSEs}
\end{table}

We tested whether using the noise model described in Section~\ref{sec:SPNA}
produced significantly different results
than when using the more complex models published in \cite{lsc+2016}. 
In that work, the SSE DE421 was used, so 
we used this SSE for a proper comparison. 
We used single-pulsar constraints on $\delm{}$ of the SSPSs using
\ttwo{}, which can use both types of noise models for single-pulsar cases
to constrain the mass error. 
This test was useful for investigating 
whether any of the pulsars had such noise properties that using
a simpler noise model would create a significant bias in the multi-pulsar,
correlated search for errors in the SSE input masses of solar-system bodies.
We did not find any statistically significant differences
between the $\delm{}$ measurements using the different noise models.
We then proceeded to compare the single-pulsar results using
\ttwo{} and the algorithm described here, implementing the noise model used in this work. 
We found the $\delm{}$ measurements to be consistent using the two different codes. 

\begin{figure*}
\begin{center}
\includegraphics[width=\textwidth]{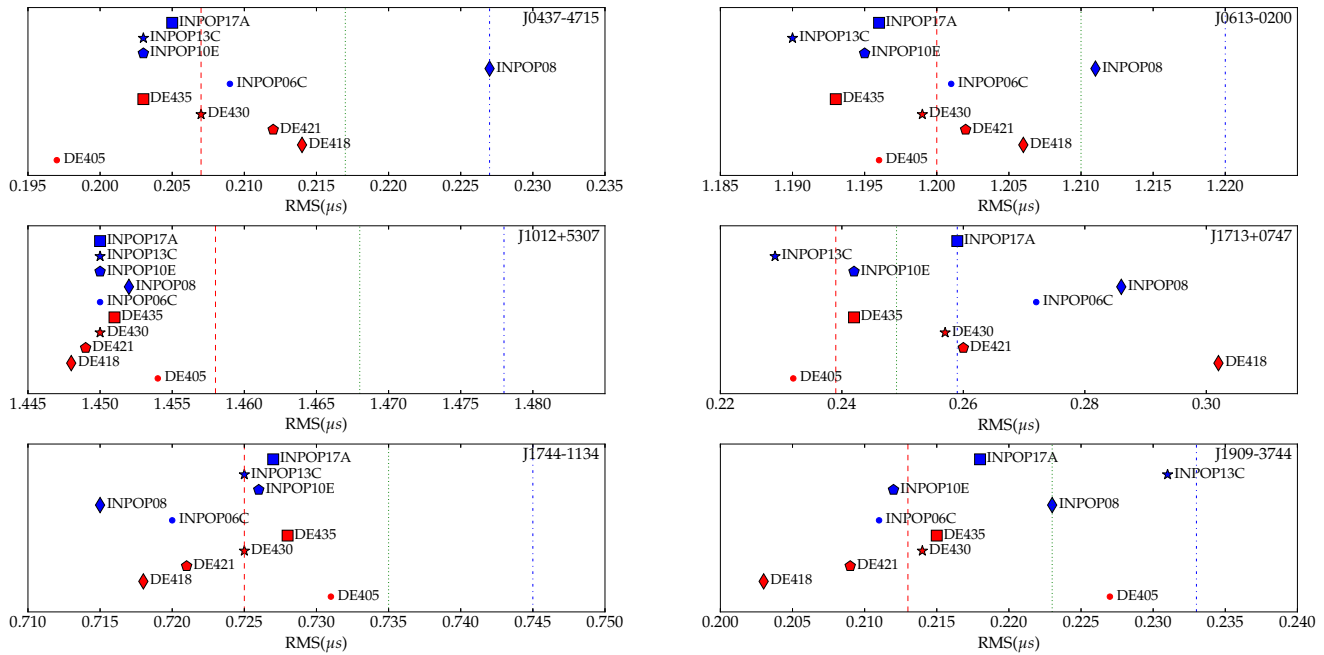}
\end{center}
\caption{The RMS of the timing residuals of the six MSPs
listed in Table~\ref{tab:MSPs}, using the ten SSEs listed in Table~\ref{tab:SSEs}. 
The dashed red, dotted green and dashed-dotted 
blue lines represent RMS values which are 
10, 20, and 30~ns larger than the smallest RMS achieved for the MSP in question.
Note that PSR~J0437$-$4715 was not included when calculating mass constraints
for solar-system bodies (see Section~\ref{sec:psrSelect}).
}
\label{fig:SSEsRMS}
\end{figure*}

We carried out a first-order examination of the effects of our choice of SSE
during the timing analysis. As the timing residuals are the primary metric
of the completeness of the timing model, we compared the
residuals' weighted root-mean-square (RMS) for each pulsar when using
different SSEs. 
The results for six MSPs (see next section for the selection of pulsars)
are summarized in Fig.~\ref{fig:SSEsRMS}. 
If we assume that the residual RMS
will be minimal for the best-performing SSE, the SSE ranking
varies for different pulsars, suggesting that the SSE performance
is dependent on the sky position. 
It is known that the differences between the pairs of SSEs have
various sky patterns, an effect that can be 
illustrated using simulated data \citep[see][]{cab2018}.

It is important to keep in mind
that SSE related residuals can be fitted out by a number of
timing parameters if they have power at those frequencies \citep*{bnr1984}. 
We are aware that this happens with parameters such as the annual term of the 
position of the pulsar and other 
astrometric parameters \citep*[see e.g.][]{mcc2013,wch+2017}.
Residual signals due to possible SSE imperfections may also be covariant with
pulsar noise parameters. 
As a result, in the absence of independent constraints
on pulsar timing parameters, the SSE ranking based on
the timing residuals RMS does not necessarily mean overall better accuracy
on the data used to construct the SSE. 
\cite{mcc2013} also demonstrated 
that the ability of the noise models included in the timing analysis
to prevent leakage of residuals in astrometric parameters depends on the
total timespan of the pulsar data set.
Therefore, a given SSE may perform differently in terms of
the residual RMS for pulsars with different time-spans, even
when their true noise properties are similar, since de-correlating
pulsar noise, SSE residuals and astrometric parameters requires sufficient data length.
Additionally, a given SSE may be over- or under-performing
by comparison to another SSE for different solar-system bodies 
when used in pulsar timing,
so the data-span can further influence the overall performance of an SSE.

While at present the differences in the RMS values 
of the residuals using different SSEs are within the noise-fluctuation levels,
it is clear that without a full account of such effects in the timing model,
cross-checking our results using various SSEs
makes studies such as the one presented in this paper more meticulous and robust.
A direct consequence of the issues discussed above is that a result
regarding the constraints on planetary masses
becomes more reliable when using
pulsars at as many sky positions as possible and with 
comparable timing precision and overall data quality, when possible.

\subsection[]{Selection of pulsars for analysis}
\label{sec:psrSelect}

The last point to consider before proceeding to the analysis 
is which pulsars to use.
Searching correlated signals with many pulsars
is a computationally intensive task.
It has thus been common practice to attempt
a ranking of the pulsars available,
in order to choose those expected to contribute 
the most to the analysis. 
The type of signal sought,
the noise characteristics of each pulsar as well as the
details of each pulsar's data quality (cadence, time-span, observing frequencies, etc)
play crucial roles in the ranking.
\begin{table}
\small
\caption{Average sensitivity to mass of UMOs in Keplerian orbits 
in four ranges of the semi-major axis, $a$, for single-pulsar cases.
The table reports the sensitivity
as the logarithms of the average of the 1$\sigma$ upper limits on the mass of UMOs
within each semi-major axis range.
The MSPs are listed in order of sensitivity (best to worse) in the interval $a\in$[5,17].
Given the data set's cadence and time-span, this is the interval where the analysis
performance is expected to impact mostly on our results.
MSPs in boldface were selected to derive the mass constraints of SSPS, ABOs and UMOs
(see discussion in main text).
}
\begin{tabular}{l c c c c}
\toprule
PSR       &    \multicolumn{4}{c}{$\log$(M/M$_{\odot}$)}   \\
Name     &       &                  &                    &    \\
(J2000)  &  $a$(AU) & $a$(AU) & $a$(AU) & $a$(AU)\\
	  &  $\in$[0.4,1.4] & $\in$[1.4,5] & $\in$[5,17] & $\in$[17,60]\\

\midrule
\textbf{J1713+0747}     & $-$9.921   & $-$9.933   & $-$8.514   & $-$5.824 \\
\textbf{J1909$-$3744}  & $-$10.040 & $-$10.436 & $-$8.317   & $-$6.077 \\
\textbf{J1744$-$1134}  & $-$9.337  &  $-$9.520  &  $-$8.200  &  $-$5.720 \\
J0437$-$4715              & $-$9.737  &  $-$9.244 &  $-$8.091  &  $-$5.861 \\
\textbf{J1012+5307}     & $-$9.113  &  $-$9.372  &  $-$7.764   & $-$5.323 \\
\textbf{J0613$-$0200}  & $-$9.323  &  $-$9.600  &  $-$7.645  &  $-$5.135 \\
\bottomrule
\end{tabular}
\label{tab:SPSCinter}
\end{table}

We made a single ranking of the pulsars
that we used for
both analysis methods described in Section~\ref{sec:Methods}
so that we are able to directly compare the results of the
analysis for modelled and unmodelled solar-system objects.
Our approach was to use the \glc{} Bayesian code described in Section~\ref{sec:bayes}
to determine the sensitivity curves of the single-pulsar data to UMO
masses. For this, we used the SSE DE421.
Table~\ref{tab:SPSCinter} shows a
breakdown of the average sensitivity in four intervals of
semi-major axis, chosen to be equal in logarithmic space.
One can see that the relative sensitivity between pulsars
can change over the semi-major axis or equivalently over the period of the
Keplerian orbit. Given that the time-spans of our pulsar data sets 
are between the orbital periods of Jupiter and Saturn while the cadence for all
pulsars is much shorter than the period of Mercury, we anticipate
that the sensitivity of the pulsars in the third 
semi-major axis interval (5$<a/	$AU$<$17)
is the most impactful to our results. 
We therefore made a priority list  
according to the average sensitivity in that interval.
Beyond the top six pulsars, the average sensitivity
drops significantly and we therefore decided not
to use more MSPs for this work.

The set we used to derive mass limits eventually
consisted of five pulsars (highlighted in Table~\ref{tab:SPSCinter}). 
Despite the fact that PSR~J0437$-$4715
is fourth in the ranking, we decided to not include it in this analysis.
This is because examination of the posterior
distribution of the orbital parameters 
from the single-pulsar Bayesian analysis for UMOs using PSR~J0437$-$4715
revealed possible systematics in the 
high-frequency regime (i.e. for small values of the semi-major axis)
which resulted in the calculated sensitivity curve
violating the analytic sensitivity curve (see \glc{} for
details on the analytic sensitivity curve).
Our analysis revealed systematics in the range of 1$<a/$AU$<$5 
which complicated the upper limit analysis, 
and worsen the UMO mass upper limits
when including this pulsar in the multi-pulsar analysis, 
in contrast to the expectation from the pulsar's noise properties and analytical sensitivity curve. 
To avoid the potential effects and complications due to these systematics,
which we reproduced using multiple SSEs,
we did not include PSR~J0437$-$4715 in constraining masses of solar-system bodies. 
This pulsar is very bright and as such has very small TOA uncertainties, but 
it is known to suffer from multiple sources of time-correlated noise \citep[see e.g.][]{lsc+2016},
which gave significant effects on our analysis exactly because of the low TOA uncertainties.
We remind the reader that, as discussed in Section~\ref{sec:SPNA}, \chm{} assumed 
their noise model for PSR~J0437$-$4715 was not precise enough around 
the orbit of Mars ($\sim~1.5$~AU).
We will focus on the results without PSR~J0437$-$4715, 
to directly compare the results of the
analysis for modelled and unmodelled solar-system objects.
Examining the exact origins of the systematics is beyond the scope of this paper
and is left for future work.

\subsection[]{Constraints on masses of known solar-system bodies}
\label{sec:ResKnown}

We used our implementation of the \chm{} method,
as described in Section~\ref{sec:DJC10} on the
five-pulsar subset of \dr{},
using the ten SSEs noted in Table~\ref{tab:SSEs}.
As explained in Section~\ref{sec:DJC10}, our analysis seeks possible errors
in the input masses, assuming that the mass error is small such that
only geometric delays of the pulse propagation 
due to errors in the estimated position of the SSB are significantly affecting the 
timing residuals. 
The SSE input values were taken directly from the header information
of the SSEs using the \textsc{calceph\_inspector} tool of the \textsc{calceph} library.

\begin{figure*}
\begin{center}$
\begin{array}{cc}
\includegraphics[width=8.6cm]{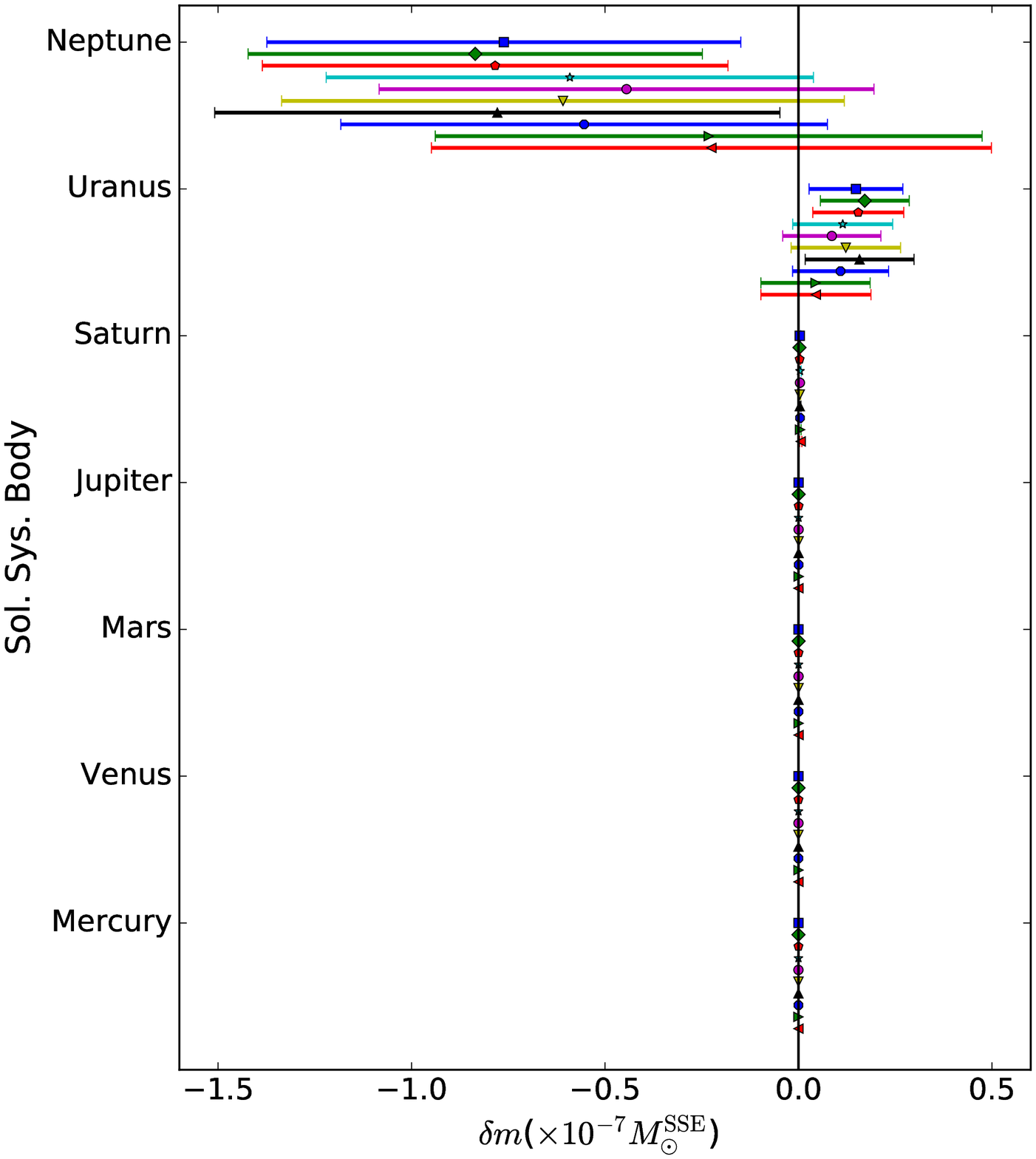} &
\includegraphics[width=8.6cm]{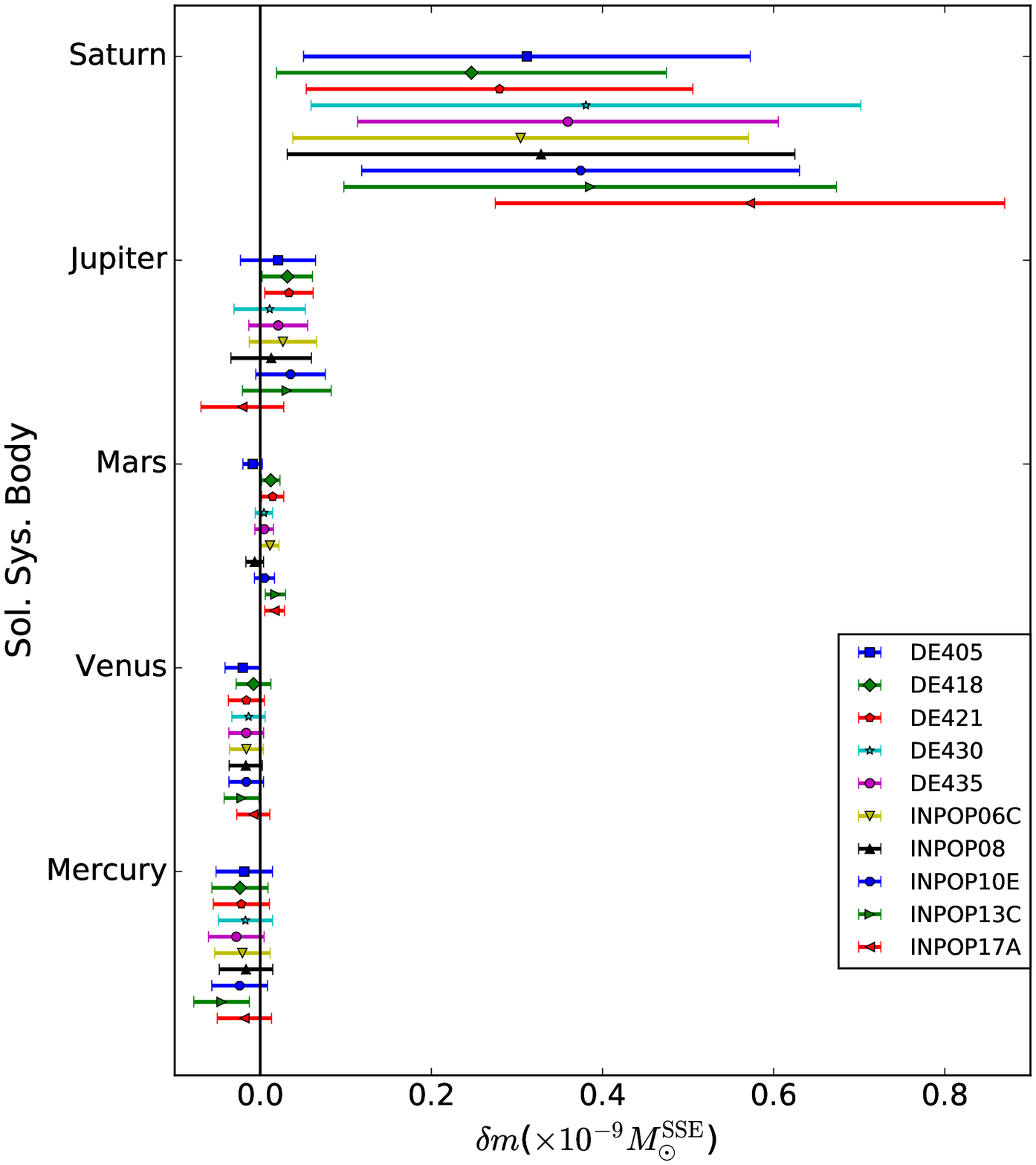} \\
\end{array}$
\caption{The derived central values and 1$\sigma$ uncertainties for errors on the 
masses of the planetary systems with respect to each SSE's input values, 
for analyses using the ten SSEs listed in Table~\ref{tab:SSEs}.
The figure on the left-hand-side includes the ice giants to emphasize the much larger
uncertainties on their derived masses.
The figure on the right-hand-side excludes the ice giants for clarity.
The $^{\textrm{SSE}}$ superscript is used to denote that each result is tied to the 
values of the Sun's gravitational mass of the given SSE
(see main text, Section~\ref{sec:ResKnown} for details).} 
\label{fig:5PSRa_Mer2Nep}
\end{center} 
\end{figure*}

Fig.~\ref{fig:5PSRa_Mer2Nep} shows the results of the analysis 
for all ten SSEs. The analysis included all planetary systems
(excluding the Earth-Moon system). 
For planets with moons we refer to the position and mass 
of the system's barycentre. The results
from the various SSEs are statistically consistent.
We also observe that despite the fact that most $\delm{}$
measurements are consistent with zero near the 1$\sigma$ level,
for each planet the central values from the various SSEs are not randomly distributed around zero
but have consistent, systematic biases, i.e. are either positive or negative.
The only exceptions are INPOP17A for Jupiter, and INPOP08 and DE405 for Mars, 
although this can be compensated by the very small values
with respect to the uncertainties. 
The most likely reason for these systematic bisases
is that at this given level of timing precision
the results are almost completely constrained
by the data, rather that from differences between SSEs within
the limits of the random noise from the measurements they use as input data.

\begin{figure}
\begin{center}
\includegraphics[width=8.5cm]{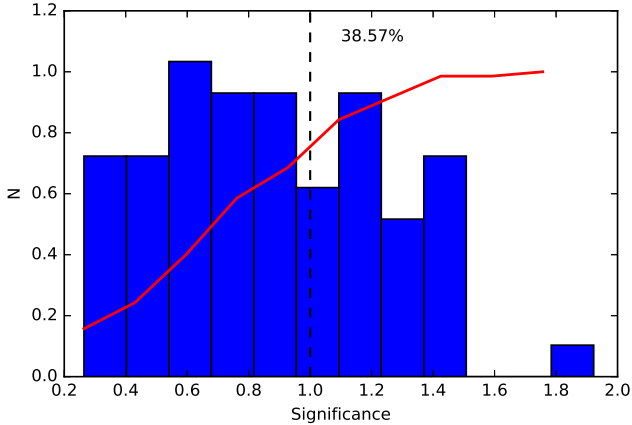}
\caption{The normalised histogram
of the distribution of the significance 
(central value divided over the 1$\sigma$ uncertainty) 
of the SSPS mass measurements. 
38.57 per cent of the cases show a significance over unity, compared
to the  31.73 per cent expected for a Gaussian distribution.
The vertical, black, dashed line indicates the significance of 1.
The red, solid line shows the corresponding cumulative distribution.} 
\label{fig:5PSRa_Mer2Nep_hist}
\end{center}
\end{figure}

Fig.~\ref{fig:5PSRa_Mer2Nep_hist} shows the distribution 
of the significance (central value divided over the 1$\sigma$ uncertainty) 
of the measurements. 
We see that 38.57 per cent of the cases (27 out of 70) show a 
measurement with a significance above 1$\sigma$.
This distribution of errors is very close 
to a Gaussian distribution(where the corresponding
percentage would be at most 31.73).
The small difference from the expected error distribution can be due to correlations
of long-orbital $\delm{}$ signals and low-frequency noise,
together with the fact that the analysis assumes symmetric 
uncertainties. That is because when a $\delm{}$
signal correlates with noise parameters, 
its probability distribution may in fact be asymmetric and the 
uncertainty would be larger on one side of the median value than the other.
Full Monte-Carlo sampling of the SSE and noise parameters
could be implemented in future work to have a better understanding 
of these correlations.


Results on Saturn and the ice giants are largely inconclusive.
The orbital periods of Uranus and Neptune 
(84 and 165 years, respectively) are more than six times longer than the data time-span,
while their masses are more than five time smaller than the mass of Saturn.
It is therefore expected, a priori, that our data set will be completely insensitive to 
any possible small errors in their masses. We include them nevertheless in our 
analysis, since the uncertainties of $\delm{}$ for these planets are
a good indication of the goodness of the uncertainties calculated 
in the presence of time-correlated noise in the pulsar data
and the sufficiency of the pulsar noise models we use.
In the presence of low-frequency noise,
if the models underestimate the noise levels, one would 
expect to see significant detections of $\delm{}$ for planets
with periods longer than the data set's time-span.
The results are as expected, with the uncertainties on
$\delm{}$ of the ice giants being orders of magnitude larger than the rest of the planets.

Using the results of the analysis, we derived the 
mass constraints of the planetary systems in the solar system
using the \dr{} and the ten SSEs employed in this study. 
The results are summarized in Table~\ref{tab:KPres}.
Since the solar gravitational parameter is a fitted quantity
in the SSEs, and is therefore different in each case,
we express all results as ratios
of the planetary gravitational parameters as derived
using a specific SSE (superscript $^{\textrm{SSE}}$) with respect to 
the nominal solar gravitational parameter, $\mathcal{GM}_{\odot}^{\textrm{N}} = 1.3271244\times10^{20}$m$^3$s$^{-2}$,
in compliance with the guidelines from the 2015 IAU Resolution B3\footnote{Available at:\\https://www.iau.org/static/resolutions/IAU2015\_English.pdf} \citep{mpt+2015}. We follow this approach in all mass constraints results we present.
While at the precision of the current data set this does not cause any differences 
in the results within the uncertainties, we nevertheless adopt this approach
to allow correct comparisons with future results.
A first observation is the consistency in the uncertainties, despite
fluctuations in the central values of  $\delm{}$. 
The \dr{} data set is sensitive to mass differences 
of a few times $10^{-11}\Ms{}$ for systems up to the 
Jovian, which constitutes a significant improvement from the
approximately $10^{-10}\Ms{}$ reported in \chm{}.
We note that for the Saturnian system, this sensitivity 
is approximately 3$\times10^{-10}\Ms{}$ while for 
the ice giants, the sensitivity drops significantly 
to approximately $10^{-8}\Ms{}$.

To evaluate our results, 
we compare them with the results from 
\chm{} and with 
the current best estimates\footnote{Up-to-date~information~at:\\ http://maia.usno.navy.mil/NSFA/NSFA\_cbe.html}
(CBEs) adopted by the International Astronomical Union (IAU) for the planet-moons systems.
The CBEs, denoted with the $^{\textrm{CBE}}$ superscript, 
are selected from the literature and are derived directly from spacecraft data.
For the comparison, we also expressed the CBE results with respect to the nominal solar gravitational parameter.
We note again, that such an approach does not change our results
within the uncertainties due to the current data precision,
but we follow this practice to allow better comparisons with future results
and follow the recommended best practices by the IAU.
Compared to \chm{}, the mass constraints 
have improved by factors
of 5.7, 8.5, 20, 6.7 and 4 for the planetary systems of
Mercury, Venus, Mars, Jupiter and Saturn, respectively.
In Table~\ref{tab:KPres2}, we also compare our
results with the IAU CBEs.
The precision of the mass constraints derived in this study for planetary systems
is lower by factors that range from of $\sim\,3$ for the case of Jupiter, up to $\sim~10^3$
for Mercury. In the case of Mercury, the large difference reflects the very significant
improvement in the planet's gravity field measurements by the MESSENGER spacecraft.
The CBEs for Mercury's gravitational mass \citep{mgg+2014} are about a factor $10^3$
more precise than the previous CBEs \citep{ace+1987}.

\begin{table*}
\centering
\caption{The mass constraints on the planetary systems derived with the \dr{}, using ten different SSEs,
expressed as ratios of their gravitational masses to that of 
the nominal solar gravitational mass.
(see main text, Section~\ref{sec:ResKnown} for details).
Numbers in brackets indicate the uncertainty in the last digit quoted.
All results are consistent at the 1$\sigma$ level.} 
\resizebox{\textwidth}{!}{\begin{tabular}{lccccccc}
\toprule
                       &       \\
Solar-system  &     \multicolumn{7}{c}{$\GMsRatioIPTANom{}$}  \\                       
Ephemeris     &   &    &    &  &  &  &   \\
             & Mercury  & Venus   &  Mars  & Jupiter & Saturn & Uranus & Neptune \\
\midrule
DE405 & 1.6600(3)$\times10^{-7}$ & 2.44782(2)$\times10^{-6}$ & 3.2271(1)$\times10^{-7}$ & 9.5479196(4)$\times10^{-4}$ & 2.858863(3)$\times10^{-4}$ & 4.368(1)$\times10^{-5}$ & 5.144(6)$\times10^{-5}$ \\
DE418 & 1.6599(3)$\times10^{-7}$ & 2.44783(2)$\times10^{-6}$ & 3.2273(1)$\times10^{-7}$ & 9.5479195(3)$\times10^{-4}$ & 2.858859(2)$\times10^{-4}$ & 4.368(1)$\times10^{-5}$ & 5.143(6)$\times10^{-5}$ \\
DE421 & 1.6599(3)$\times10^{-7}$ & 2.44782(2)$\times10^{-6}$ & 3.2273(1)$\times10^{-7}$ & 9.5479195(3)$\times10^{-4}$ & 2.858860(2)$\times10^{-4}$ & 4.368(1)$\times10^{-5}$ & 5.144(6)$\times10^{-5}$ \\
DE430 & 1.6599(3)$\times10^{-7}$ & 2.44782(2)$\times10^{-6}$ & 3.2272(1)$\times10^{-7}$ & 9.5479193(4)$\times10^{-4}$ & 2.858861(3)$\times10^{-4}$ & 4.367(1)$\times10^{-5}$ & 5.145(6)$\times10^{-5}$ \\
DE435 & 1.6598(3)$\times10^{-7}$ & 2.44782(2)$\times10^{-6}$ & 3.2272(1)$\times10^{-7}$ & 9.5479193(3)$\times10^{-4}$ & 2.858860(2)$\times10^{-4}$ & 4.367(1)$\times10^{-5}$ & 5.147(6)$\times10^{-5}$ \\
INPOP06C & 1.6599(3)$\times10^{-7}$ & 2.44782(2)$\times10^{-6}$ & 3.2273(1)$\times10^{-7}$ & 9.5479194(4)$\times10^{-4}$ & 2.858863(3)$\times10^{-4}$ & 4.367(1)$\times10^{-5}$ & 5.145(7)$\times10^{-5}$ \\
INPOP08 & 1.6600(3)$\times10^{-7}$ & 2.44782(2)$\times10^{-6}$ & 3.2271(1)$\times10^{-7}$ & 9.5479193(5)$\times10^{-4}$ & 2.858863(3)$\times10^{-4}$ & 4.368(1)$\times10^{-5}$ & 5.144(7)$\times10^{-5}$ \\
INPOP10E & 1.6599(3)$\times10^{-7}$ & 2.44782(2)$\times10^{-6}$ & 3.2272(1)$\times10^{-7}$ & 9.5479193(4)$\times10^{-4}$ & 2.858860(3)$\times10^{-4}$ & 4.367(1)$\times10^{-5}$ & 5.146(6)$\times10^{-5}$ \\
INPOP13C & 1.6597(3)$\times10^{-7}$ & 2.44782(2)$\times10^{-6}$ & 3.2273(1)$\times10^{-7}$ & 9.5479193(5)$\times10^{-4}$ & 2.858861(3)$\times10^{-4}$ & 4.367(1)$\times10^{-5}$ & 5.149(7)$\times10^{-5}$ \\
INPOP17A & 1.6599(3)$\times10^{-7}$ & 2.44783(2)$\times10^{-6}$ & 3.2273(1)$\times10^{-7}$ & 9.5479189(5)$\times10^{-4}$ & 2.858862(3)$\times10^{-4}$ & 4.367(1)$\times10^{-5}$ & 5.149(7)$\times10^{-5}$ \\
\\
\bottomrule
\end{tabular}}
\label{tab:KPres}
\end{table*}

\begin{table}
\small
\caption{Comparison between the mass constraints on the planetary systems
from this work (IPTA1), the \chm{} results and the CBEs adopted by the IAU.
Numbers in brackets indicate the uncertainty in the last digit quoted.
the different results are expressed in terms
of the nominal solar gravitational mass (see main text, Section~\ref{sec:ResKnown} for details).
The sensitivity of the methods can be compared via the ratio
of their uncertainties ($\sigma$). For IPTA values, we used the
case with the highest uncertainty for each planetary system. 
Where multiple SSE cases gave the same uncertainty,
we note the mass constraint derived with the most recent SSE.
The IPTA and IAU have the most 
comparable mass uncertainties in the case of the Jovian system.
The largest difference in the case of Mercury.
}
\resizebox{.48\textwidth}{!}{\begin{tabular}{>{\LARGE}l>{\LARGE}r>{\LARGE}c>{\LARGE}r>{\LARGE}c}
\toprule
Planetary &  \multicolumn{1}{c}{\LARGE{$\GMsRatioIPTANom{}$}}          &    \LARGE{$\CHMIPTAratio$}&  \multicolumn{1}{c}{\LARGE{$\GMsRatioIAUNom$}}        & \LARGE{$\IPTAIAUratio{}$}\\
System    & \multicolumn{1}{c}{}     &  &  \multicolumn{1}{c}{}& \\
\midrule
Mercury &  1.6599(3)$\times10^{-7}$ & 5.5 &1.66012099(6)$\times10^{-7}$ & 5.3$\times10^{3}$\\
Venus   &   2.44783(2)$\times10^{-6}$ & 8.5  & 2.44783824(4)$\times10^{-6}$ & 50.0\\
Mars     &  3.2273(1)$\times10^{-7}$ & 20 &  3.2271560(2)$\times10^{-7}$ & 500.0\\	
Jupiter  &   9.5479189(5)$\times10^{-4}$ & 6.7 & 9.54791898(16)$\times10^{-4}$ & 3.13 \\
Saturn  &   2.858863(3)$\times10^{-4}$ & 4.0 & 2.85885670(8)$\times10^{-4}$ & 37.5\\
Uranus  &   4.367(1)$\times10^{-5}$ &n.a. &4.366249(3)$\times10^{-5}$ & 333.3\\
Neptune&   5.149(7)$\times10^{-5}$ &n.a. &5.151383(8)$\times10^{-5}$ & 875.0\\
\bottomrule
\end{tabular}}
\label{tab:KPres2}
\end{table}

\subsubsection[]{Asteroid-belt objects}
\label{sec:ResKnownAst}

The main asteroid belt hosts small bodies with masses that reach up to
order $10^{-10}\,\Ms{}$. With the \dr{} having sensitivity to mass errors of the 
order $10^{-11}\,\Ms{} - 10^{-10}\,\Ms{}$ between the orbits of Mars and Jupiter
(see also next section), it is logical to attempt constraining the masses of the largest
bodies of the main belt. This is the first time that PTA data 
are used to derive mass constraints on ABOs.
As our data are only beginning to be sensitive to ABO masses,
in this work we perform a pilot study and use only one SSE. 
Future work with more sensitive data can focus more on comparisons
between the pulsar-timing constraints on ABO mass using different SSEs.
We employed the SSE DE435 together with additional, high-precision
positional data for the ABOs from the
\emph{New Horizons SPICE Data Archive}\footnote{https://ssd.jpl.nasa.gov/x/spk.html},
which were used for the New Horizons spacecraft mission.
These auxiliary data are provided by JPL in the \textsc{spice} kernel format
and for this reason, for this application we use the \textsc{spice} library and tools.

\begin{figure}
\begin{center}
\includegraphics[width=8.8cm]{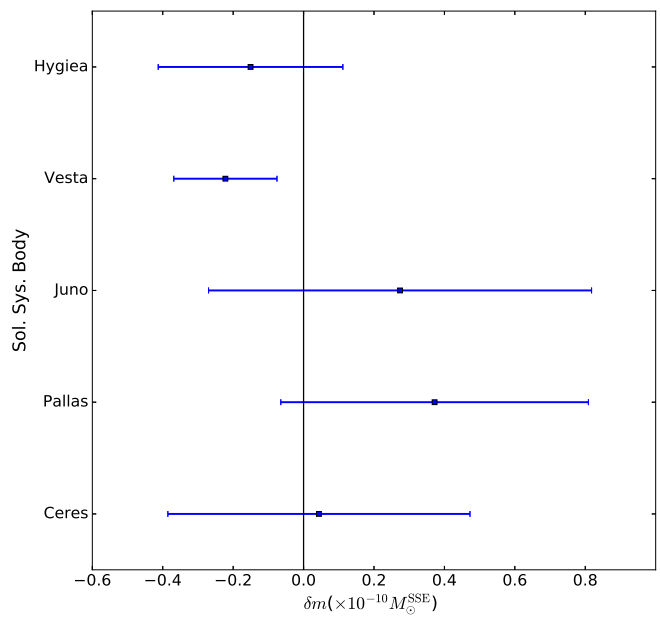}
\caption{The derived central values and 1$\sigma$ uncertainties for errors on the 
masses of five massive asteroid-belt objects with respect to the SSE's input values, 
for an analysis using the DE435 SSE
and updated
high-precision positional data from the 
\emph{New~Horizons~SPICE~Data~Archive}.
The $^{\textrm{SSE}}$ superscript is used to denote that each result is tied to the 
values of the Sun's gravitational mass of the given SSE
(see main text, Section~\ref{sec:ResKnown} for details).} 
\label{fig:5PSRa_Cer2Hyg}
\end{center} 
\end{figure}

The $\delm{}$ measurements are shown in Fig.~\ref{fig:5PSRa_Cer2Hyg} and
Table~\ref{tab:KAres} presents the mass constraints derived.
We produced IPTA mass constraints on the three ABOs included in 
the IAU body constants, namely the dwarf planet Ceres
and the asteroids Pallas and Vesta and additionally for
another two large asteroids, Juno and Hygiea. 
For both Ceres and Pallas,
the IPTA mass constraint is only slightly over an order of magnitude
larger than the IAU CBEs. On the other hand,
the IPTA precision on the mass of Vesta is five orders of magnitude worse.
This is because of a very precise new determination of the asteroid's
mass, orbital and orientation parameters by \cite{kap+2014},
which increased the precision of the mass measurement by 
a factor 10$^{5}$ from the previous best estimate. 
This was achieved by measurements made 
with radiometric tracking and optical
data from the Dawn spacecraft \citep{rr2011}. 
The Dawn space mission was specifically designed to
send the spacecraft in orbit around Ceres and Vesta for detailed studies.
We note that although not yet adopted by the IAU, a publication
has recently appeared presenting results for Ceres by the Dawn
mission, which has also improved the precision of its mass measurement
by a factor of 100 \citep{kpv+2018}. For the asteroids Juno and Hygiea the 
uncertainty is equal or higher than the mass constraint
and therefore we can only assume upper limits of 
9$\times10^{-11}\,M_{\odot}$ and 6$\times10^{-11}\,M_{\odot}$ on their masses,
respectively, at the 68 per cent confidence level.
\begin{table}
\caption{
Comparisons of the
mass constraints for 
the five most massive ABOs derived in this work with
the IAU CBEs. 
When IAU CBEs are unavailable (noted with $^{\star}$ superscript),
we use the values from \citet{car2012}.
The IPTA masses were derived
using the SSE DE435 and updated
high-precision positional data from the 
\emph{New~Horizons~SPICE~Data~Archive}.
For the comparison, the different results are expressed in terms
of the nominal solar gravitational mass (see main text, Section~\ref{sec:ResKnown} for details).
Numbers in brackets indicate the uncertainty in the last digit quoted.
}
\resizebox{.48\textwidth}{!}{\begin{tabular}{>{\LARGE}l>{\LARGE}c>{\LARGE}c>{\LARGE}c>{\LARGE}c}
\toprule
Name       & Minor Planet & $\GMsRatioIPTANom{}$  & $\GMsRatioIAUNom$ & $\IPTAIAUratio$\\
       & Category                &  &  & \\
\midrule
1 Ceres & Dwarf Planet & 4.8(4)$\times10^{-10}$ & 4.72(3)$\times10^{-10}$ & 13.3\\
2 Pallas & Asteroid        & 1.4(4)$\times10^{-10}$ & 1.03(3)$\times10^{-10}$ & 13.3\\
3 Juno${^\star}$ & Asteroid          &4(5)$\times10^{-11}$& 1.37(1)$\times10^{-11}$ & 500\\
4 Vesta & Asteroid         & 1.1(1)$\times10^{-10}$ & 1.3026846(9)$\times10^{-10}$ & 1.1$\times10^{5}$\\
10 Hygiea${^\star}$ & Asteroid     & 3(3)$\times10^{-11}$&4.3(3)$\times10^{-11}$ & 100\\
\bottomrule
\end{tabular}}
\label{tab:KAres}
\end{table}
 
\subsection[]{Constraints on masses of UMOs}
\label{sec:ResUnKnown}
\begin{figure}
\begin{center}$
\begin{array}{cc}
\includegraphics[width=8.5cm]{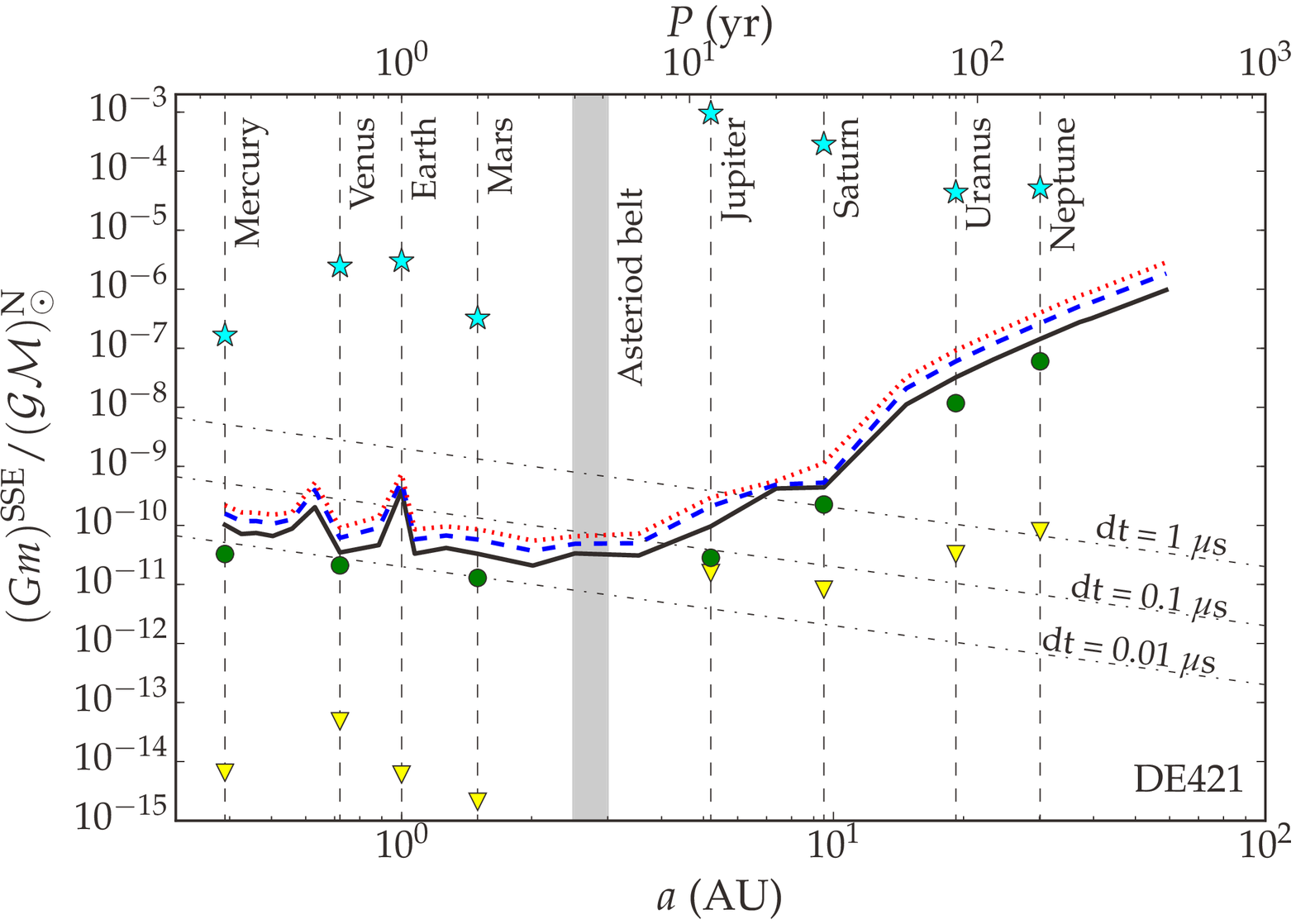} \\
\includegraphics[width=8.5cm]{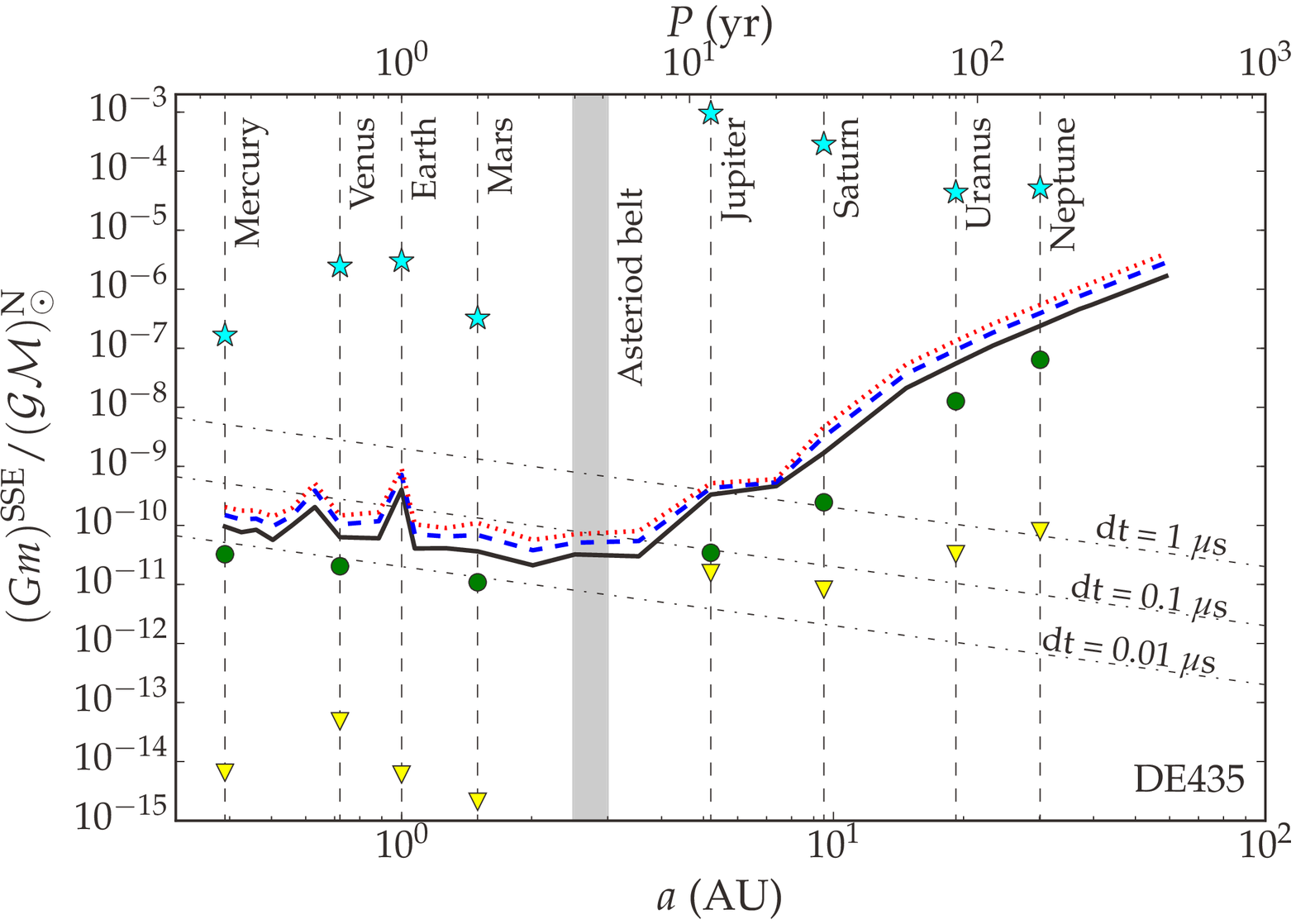} \\
\includegraphics[width=8.5cm]{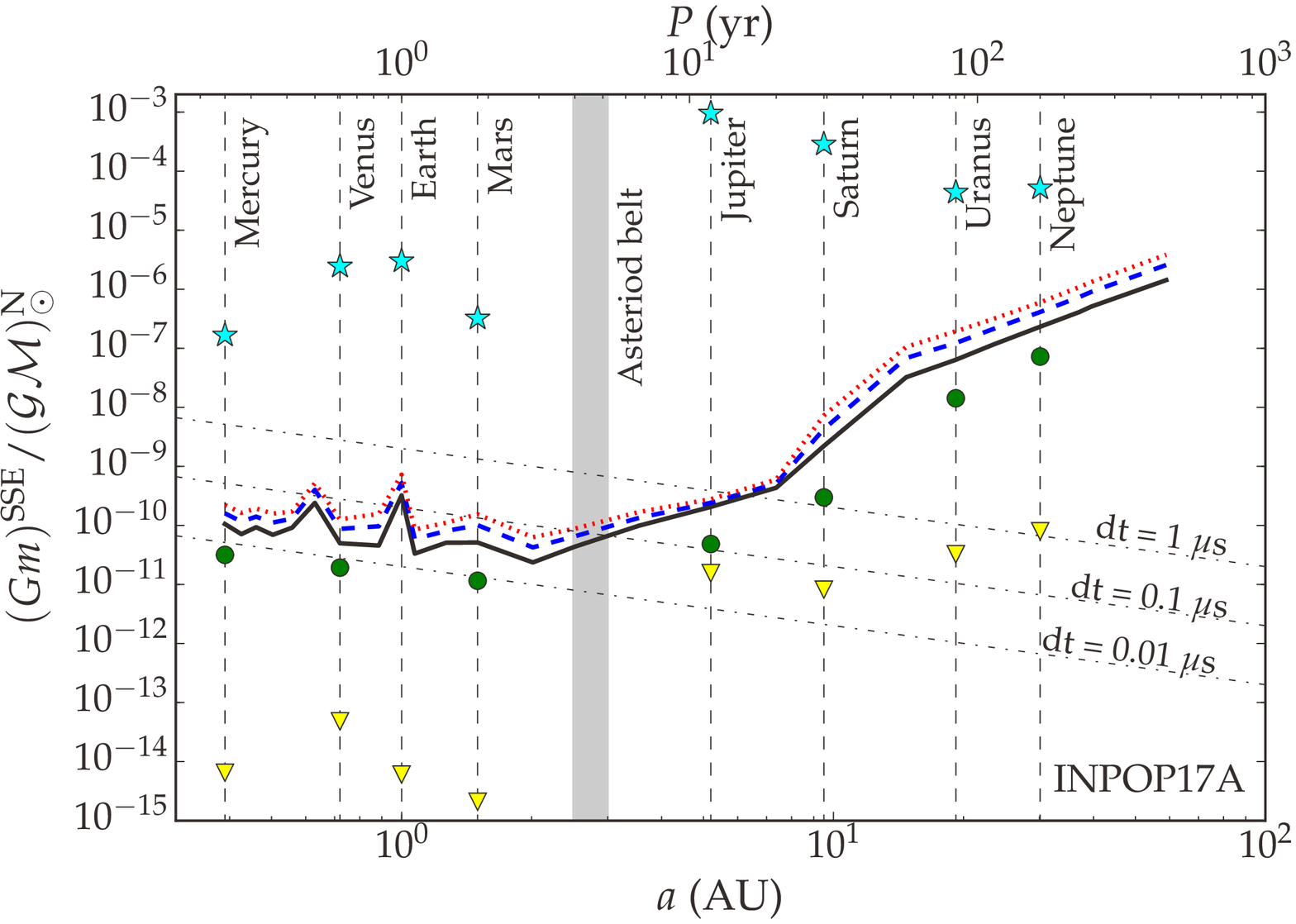} \\
\end{array}$
\caption{Sensitivity curves for unmodelled masses in Keplerian orbits
for the \dr{}, using three different SSEs. From top to bottom, the figures 
show the 1$\sigma$ (solid, black), 2$\sigma$ (dashed, blue) 
and 3$\sigma$ (dotted, red)
upper limits (corresponding to the 68, 95 and 99.7 per cent credible intervals of the posterior distribution) 
for the mass for the DE421, DE435 and INPOP17A SSEs, respectively.
The dot-dashed lines show the expected amplitude of the residuals
induced by a given mass in Keplerian orbit at any given semi-major axis value.
The grey shaded region shows
the position of the asteroid belt.
The cyan stars indicate the official IAU masses for the planetary systems. 
The corresponding IAU uncertainties and uncertainties 
from our analysis of known SSPSs
on the planetary masses are plotted as yellow 
triangles and green circles, respectively, for comparison.
} 
\label{fig:5PSRa_USSBDs_A}
\end{center} 
\end{figure}
\begin{figure}
\begin{center}
\includegraphics[width=8.5cm]{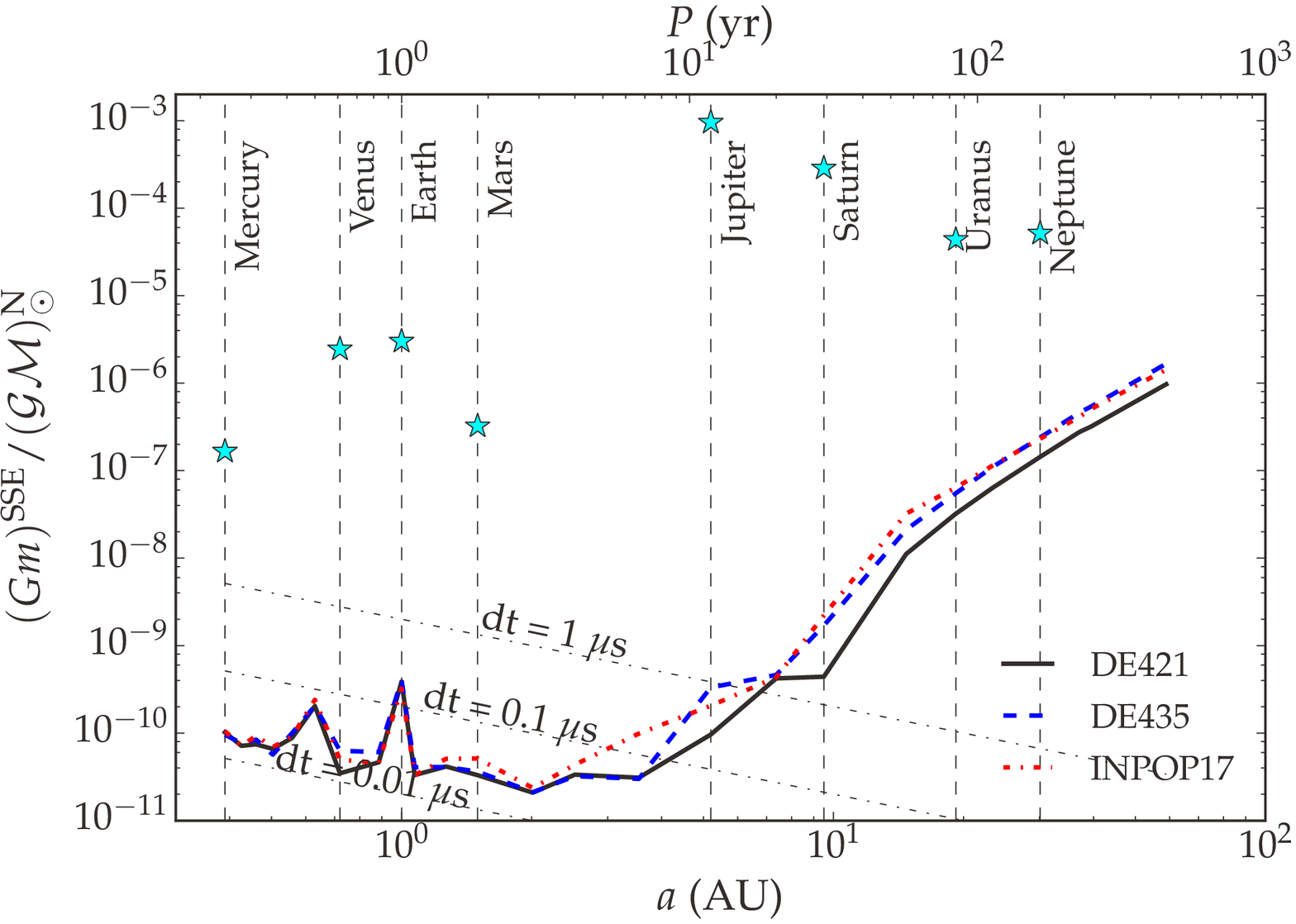} \\
\caption{A comparison of the 1$\sigma$ upper limits
for the three used SSEs presented in Fig.~\ref{fig:5PSRa_USSBDs_A}. 
} 
\label{fig:5PSRa_USSBDs_B}
\end{center} 
\end{figure}
We used the same five-pulsar list as in the analysis for the 
SSPSs and ABOs in the previous section and
employed the method outlined in Section~\ref{sec:bayes} to conduct
the Bayesian analysis to search for UMOs.
This is the first time that such an analysis has been conducted using
real PTA data. 
Given the very high consistency
in the results produced using the ten
SSEs in the previous section, we opted
to focus on three SSEs, namely DE421, DE435 and INPOP17A.
The first was chosen for comparison reasons, since it is
the SSE used in \chm{} and the \dr{} data release and 
noise-analysis papers \citep{vlh+2016,lsc+2016},
while the other two were chosen because they are the 
latest from each SSE family among those used in this study.
Table~\ref{tab:priors} gives an overview of the types of prior probability distributions 
used for the sampled parameters, as well as the ranges of their values.

\begin{table}
\caption{Prior types and ranges for the 
Bayesian analysis to constrain the masses of UMOs.
Two sets of priors are shown, one for the blind search of UMOs
and one for the mass upper limit analysis (See discussion on priors in Section~\ref{sec:bayes}).}
\begin{tabular}{lcc}
\toprule
Parameter  & \multicolumn{2}{c}{Prior range}\\
                  & Blind   &  Upper-limit\\
                  & search & analysis \\
\midrule
$m$ ($M_{\odot}$) & log-uniform in $\lbrack10^{-25},10^{-5}\rbrack$& uniform in $\lbrack0,10^{-5}\rbrack$ \\
$a$ (AU) & log-uniform in $\lbrack0.1, 10\rbrack$& fixed in $\lbrack0.4,60\rbrack$ \\
$e$ & uniform in $\lbrack0,0.99\rbrack$ &  uniform in $\lbrack0,0.99\rbrack$ \\
$\Omega$ & uniform in \lbrack0,2$\uppi{}$\rbrack & uniform in \lbrack0,2$\uppi{}$\rbrack  \\
$i$ & uniform in \lbrack0,$\uppi{}$\rbrack & uniform in \lbrack0,$\uppi{}$\rbrack\\
$\omega$ & uniform in \lbrack0,2$\uppi{}$\rbrack & uniform in \lbrack0,2$\uppi{}$\rbrack\\
$\phi_{0}$ & uniform in \lbrack0,2$\uppi{}$\rbrack & uniform in \lbrack0,2$\uppi{}$\rbrack\\
\bottomrule
\end{tabular}
\label{tab:priors}
\end{table}

We performed a blind orbital analysis, i.e.
we fully searched over the UMO mass and orbital parameters. 
Our analysis was restricted 
to circular and eccentric orbits. 
For all three SSE cases we derived a non-detection result,
and produced the mass sensitivity curves,
which we present in Fig.~\ref{fig:5PSRa_USSBDs_A}.
In Fig.~\ref{fig:5PSRa_USSBDs_B} we overplot the results
from the three cases for direct visual comparison.
The results show how the relative sensitivity of the data
at various distances from the SSB changes
when using different SSEs.
While for semi-major axis values up to the orbit of Mars the three SSEs are 
in very close agreement, for wider orbits the results
are less consistent, with DE421 showing overall higher sensitivity,
i.e. giving the lowest upper limits. DE435 and INPOP17A
show their biggest differences in the 
semi-major axis range 4~--~8\,AU,
i.e. in the asteroid belt, and around the orbit of Jupiter,
and become fully consistent for distances beyond 20\,AU.
Table~\ref{tab:UPres} presents the upper limits
on the mass of UMOs at selected values
of the semi-major axis, for the three SSEs used. 

Direct comparison to the results for known SSPSs
at the same semi-major axis values
is only approximate, since this analysis assumes unperturbed
Keplerian orbits, in contrast to the analysis in the 
previous section which follows the exact orbits based on observations.
It is nevertheless useful to make the comparison as a cross-check,
since the much larger degrees of freedom in the search for UMOs
should always result in worse sensitivity by comparison to 
that of known bodies for the same same-major axis values.
This is indeed the case in our analysis, 
with the upper limits from the blind search
being $\sim$2 to 14 times higher.
One could also extend the upper-limit analysis to wider orbits
in order to retrieve, for example, an upper limit on the mass of Planet Nine.
In \glc{}, the results using simulated data show that the precision of the \dr{} is
not sufficient to give informative constraints on the mass of Planet Nine.
We therefore did not attempt this, but reserve such effort for future work.

As discussed in \glc{}, the results from this type of analysis
directly provide upper limits on the presence of any type
of massive objects in orbit around the SSB. As such,
our results are also applicable to more exotic objects
such as dark matter clumps \citep{lb2005}
or cosmic strings \citep{bos2014}.
For distances above 2\,AU from the SSB
(where the sensitivity is maximum) 
we can exclude (with a 68 per cent confidence level)
the presence of dark matter 
clumps (in eccentric, Keplerian orbits) with masses up to 1.2$\times10^{-11}\,\Ms{}$.
For distances up to Saturn's orbit ($\approx$\,9.6\,AU), 
the upper limits range between
4$\times10^{-10}$ $-$ 2$\times10^{-9}\,\Ms{}$ 
(depending on the used SSE). 
For comparison, we note that \cite{ppA2013} and \cite{pAp2013}
present upper limit of 1.7$\times10^{-10}\,\Ms{}$
for the dark matter mass in the sphere
within Saturn's orbit, using independent data and methodology.
Their approach searches for perturbations on the 
orbital motion of planets due to the acceleration by an 
assumed dark matter distribution in the interplanetary space.
This comparison is only indicative, since that work 
assumes that dark matter has a continuous distribution that is 
spherically symmetric relative to the Sun, with a fixed central density and exponential 
drop with increased distance \cite{ppA2013,pAp2013}. Other density distributions are also
discussed, but none of those models assumes clumps as we did in this study.
We note that \cite{ppA2013} employ an SSE independent of the ones used in this study.
Specifically, they use the EPM2011 \citep{pit2013} which is published
by the Institute of Applied Astronomy of the Russian Academy of Sciences.

\begin{table}
\caption{Derived upper limits for the mass of UMOs in Keplerian orbits
around the SSB. The upper limits quoted correspond to the 95 per cent credible intervals of the 
posterior distributions.}
\resizebox{.80\width}{!}{\begin{tabular}{lccc}
\toprule
  &          \\
 Semi-major         &    \multicolumn{3}{c}{$\GMsRatioIPTAumoNom$}      \\
Axis  &   &    &     \\
(AU)             & INPOP17A  & DE435   &  DE421 \\
\midrule
0.5 &1.14815362$\times10^{-10}$ &   1.00000000$\times10^{-10}$ &   1.07151931$\times10^{-10}$ \\
1.4 & 7.24435960$\times10^{-11}$ &   5.88843655$\times10^{-11}$ &   6.02559586$\times10^{-11}$ \\
5.0 & 2.23872114$\times10^{-10}$ &   1.28824955$\times10^{-10}$ &   1.20226443$\times10^{-10}$ \\
10 & 2.95120923$\times10^{-09}$ &   2.39883292$\times10^{-09}$ &   1.86208714$\times10^{-09}$ \\
17 & 9.12010839$\times10^{-08}$ &   5.62341325$\times10^{-08}$ &   3.31131121$\times10^{-08}$ \\
60 & 2.57039578$\times10^{-06}$ &   2.81838293$\times10^{-06}$ &   1.86208714$\times10^{-06}$ \\
\bottomrule
\end{tabular}}
\label{tab:UPres}
\end{table}

\section{Discussion and Conclusions}
\label{sec:Disc}

In the work described in this paper
we have employed previously published methods
on a subset of the first IPTA data release in order
to constrain the masses of solar-system bodies 
using ten different SSEs,
five from IMCCE and five from JPL. 
Using a new computational implementation
of the method first described in \chm{},
we have derived new
mass constraints for the SSPSs, 
which were found to be statistically consistent using all ten SSEs.
While the biases from the SSE reference values
appear consistent for each SSPS, the results appear 
to be dominated by data noise. Within the uncertainties,
our results are in agreement with the CBEs from the IAU
which overall have significantly lower uncertainties.
For the first time, PTA data were also used to significantly
constrain the masses of the most massive ABOs.
A Bayesian method from \glc{} was also employed
for the first time on real data to provide generic
sensitivity limits on the mass of UMOs in the solar system using pulsar timing.

The new mass constraints
on all planetary systems show improvements
of factors 4 to 20 from the last work that used the same method, namely \chm{},
emphasizing the fact that increasing the precision, cadence, frequency coverage and
time-span of the pulsar-timing data allows for constant
improvements of PTA sensitivity to potential errors in SSEs.
As such, the IPTA greatly serves this research since the
combination of independent data sets 
improves the data overall in all these aspects. 
As noted in \cite{cab2018},
the use of the IPTA combined data 
improved the sensitivity to 
planetary masses by factors up to $\sim~4$ 
by comparison to only using EPTA data, when using the same pulsars in both cases.
Additionally, the IPTA combined data set also allowed
more options with regards to choosing MSPs for the analyses
and this study has benefited from using a larger and different sample of pulsars than \chm{}.
We note that constraining planetary masses with pulsar-timing
data helps us cross-check the data quality and pulsar noise models
using information on physical properties 
that are measured completely independently.
Large deviations from
the SSE's reference masses or 
unexplained signals present only in  
one pulsar's data, which are not detected
with multi-pulsar searches for correlated signals, can indicate 
insufficiencies of the noise models
or possible systematics in the data of a given pulsar.

In this paper we have additionally demonstrated with real data
the ability of algorithms that search for UMOs of any type in the solar system,
to provide generic mass sensitivity curves using pulsar timing. 
While with certain limitations,
the \glc{} code applied in this paper highlights
differences between SSEs. 
As we saw in Section~\ref{sec:ResUnKnown},
the main differences in UMO-mass sensitivity curves between 
DE421, DE435 and INPOP17A appear in the asteroid belt and around
Jupiter. While the details of the differences between SSEs are 
beyond the scope of this study, we note that these 
results may be due to changes in the way that 
ABO masses and their perturbations on each other and
on Mars are estimated,
as well as recent updates in the positional data of 
Mars, Jupiter and Saturn 
that IMCCE and JPL have been implementing \citep[e.g.][]{vfg+2017,fwb+2014}.
For example, when estimating perturbations
of the ABOs on the orbit of Mars, 
both DE421 and DE435 used data for the 343 ABOs 
identified to be dominant.
However, while in DE421 only eleven ABO masses were individually calculated
(for the rest either the initial values were kept fixed 
or values were fixed to approximate values derived densities
assumed per taxonomic class),
for DE435 (also the case for DE430) the individual masses were
calculated for all 343 ABOs. 
Future work with more precise data sets
could focus more on the effects of such difference on
pulsar timing and applications.

As noted in \chm{}, since PTAs are sensitive to the 
total mass of the SSPSs, if PTAs in the future
measure differences in the masses with statistical significance,
those differences may reflect 
differences in the masses or total number of 
moons taken into account when estimating the position
of the planet-moons barycentre and total mass estimations.
Although the \glc{} algorithm
is not directly applicable to bodies in orbit around major planets,
the differences in the sensitivity curves around
semi-major axis values close to planetary orbits
may still be associated with such errors.
The mass of UMOs at these semi-major axis values,
may also reflect differences in the SSEs regarding the
positional data of the planets and moons, 
since with the applied
methodology such effects could potentially be absorbed 
by the UMO mass parameter.
These results underline the potential of pulsar timing and PTA research
to also provide feedback and independent checks to groups developing SSEs, 
and add information in the future for SSE development.

While at the precision that the \dr{}
can probe the masses of SSPSs we have confirmed that the SSEs
give consistent results,
we found that for any given MSP
the timing residuals resulting from using different SSEs can vary at different levels.
For individual pulsars we have noted that the differences between 
the RMS deviations of residuals formed using various SSEs (see Fig.~\ref{fig:SSEsRMS})
were up to $\approx65$\,ns (which corresponds to relative differences of up to 22 per cent).
The consistency in the SSPS masses from the various SSEs
in the presence of the timing-residual differences means that, to a large extent,
the levels of noise are such that the $\delm{}$ uncertainties compensate
for these residual differences.
These results, however, motivate further research into the role 
of SSEs in pulsar timing models, beyond the effects on the mass constraints
of solar-system bodies.

It is worth noting that SSEs are regularly being updated with new data
(DE436 is also available and the reader can see it applied to PTA data in \cite{abb+2018})
and the IAU regularly evaluates new data and updated the published CBEs.
As we have seen in Section~\ref{sec:ResKnownAst} for the cases of Mercury and Vesta, 
new data from space missions can indeed give at times dramatic improvements in the
measurements of masses and other physical properties of solar-system bodies.
As such we will regularly have to check the impact 
of SSE updates on pulsar-timing and PTA applications and compare our results to updated CBEs. 
Pulsar-timing results, on the other hand, are also able to show 
strong improvements over time. 
As was noted already in \chm{},
pulsar timing has the benefit of being able to improve on mass constraints with
more accumulation of data, even if the data quality remains constant.
The example of Saturn is important to highlight,
since within the next ten years our data sets will be long enough to fully
sample its orbit, which has a period of 29.5~yr. 
This will allow us to fully de-correlate signals from Saturn from those
of timing and noise parameters and the uncertainties of the Saturnian $\delm{}$
will be very significantly reduced.
General predictions on expected improvements 
in probing parameters of solar-system bodies with future 
instruments can be made using generic sensitivity curves for UMOs.
In \glc{}, it was shown that regular observations of  20-40 MSPs with future radio telescopes,
which may achieve timing residuals of average RMS levels of order $\sim~100$\,ns for 20\,yr, can 
potentially improve our mass constraints on the Jovian system by another
two orders of magnitude, at levels below the current constraints by space missions.
Therefore, despite the anticipated improvements in the CBE mass values 
of the Jovian system as a result of the analysis of data by the JUNO and JUICE space missions,
our predictions are indicative of the potential for interesting
results that can be produced in the future 
with respect to solar-system studies using PTA data. 

Finally, we comment on the relation between research into 
SSEs, solar-system studies and GW searches with PTAs. 
Results of studies such as the present
can give hints on which GW frequencies one can expect
most of the differences in the limits by PTAs when using different SSEs.
For example, the results shown in Fig.~\ref{fig:5PSRa_USSBDs_B}
suggests that searches for GWs using DE421 and DE435 would be
mostly affected by SSE choice around GW frequencies $\sim2-8\,$nHz.
The sensitivity of PTAs to the 
dimensionless strain of stochastic GWBs is currently of the order of 
10$^{-15}$ at reference frequency 1\,yr$^{-1}$
\citep[e.g.][]{vlh+2016}.
As an example, let us consider the case of a GWB
formed by the superposition of GWs for a large number of
GW-driven supermassive black-hole binaries,
which have a dimensionless strain that scales
with the GWB frequency as $f_{\textrm{\tiny{gwb}}}^{-2/3}$,  \citep[e.g.][]{ses2013}.
The RMS of such a GWB signal would then be at
levels $\lesssim200$\,ns. Depending on the pulsar position,
differences between SSEs are shown to vary between $\sim~$15~--~450\,ns \citep{cab2018}.
It is therefore rational to anticipate that once other sources of noise,
such as IISM related chromatic noise, are mitigated,
the GWB searches and limits will begin to depend more clearly on
the choice of SSE. 
This is indeed the case and already
GWB upper limits and detection statistics are 
being affected by the choice of SSEs, 
which leads to the need of introducing SSE-related parameters 
in the model in order to mitigate such effects \citep{abb+2018}.
Vice versa, one expects the presence of a GWB
to influence our results when trying to constrain planetary masses with PTA data.
As discussed in Section~\ref{sec:intro}, with more MSPs of high data precision, 
a GWB and a $\delm{}$ signal should be
distinguishable on the basis of their different angular correlation. Further work is
underway to understand the impact of SSE selection to the timing and noise models
and to bridge the systematics of each SSE to a given analysis. 

Early attempts to do this included non-physical, generic error vectors
of the position of the SSB \citep{dhy+2013,thk+2016}, 
which was implemented with real EPTA data in \cite{ltm+2015M}. The latter
study only used one SSE, since there was no evidence of the GWB-strain
upper limits being influenced by adding this SSE component in the model.
In this case, the effects of the ephemeris error added in the Bayesian model
and the relevant parameters were simultaneously sampled with GWB parameters.
Further work in this direction was demonstrated in \cite{tlb+2017}. 
Using simulated data, they recovered the signal induced by an error in the
SSE's input Jupiter mass, correctly estimated the value of the error in the mass,
and were able to distinguish the mass-error signal from a GWB signal.
More recently, \cite{abb+2018} implemented a physical SSE-perturbation
model that allows a combination of coordinate-frame drifts, gas-giant mass perturbations 
(as in this paper), and Jupiter orbital-element perturbations. 
Their findings indicated that upper limits and signal-vs-noise odds ratios 
for a GWB can vary significantly depending on the choice of SSE.
The new model led to identical SSE-marginalized GWB statistics, 
regardless of the initial SSE model (both JPL and IMCEE models were used).
Both this work and \cite{abb+2018} the models were limited in the use of 
parameters describing linear mass-perturbation effects
on the TOAs, but further components will be employed in the future using the upcoming new
IPTA data releases.

\section*{Acknowledgments}\label{}
 
Part of this work is based on observations with the 100-m telescope of the Max-Planck-Institut f{\"u}r 
Radioastronomie (MPIfR) at Effelsberg. The Nan\c cay Radio Observatory is operated by the Paris Observatory, 
associated to the French Centre National de la Recherche Scientifique (CNRS). Pulsar research at 
the Jodrell Bank Centre for Astrophysics and the observations using the Lovell Telescope are supported by a 
consolidated grant from the STFC in the UK. The Westerbork Synthesis Radio Telescope is operated by the 
Netherlands Institute for Radio Astronomy (ASTRON) with support from The Netherlands Foundation for Scientific 
Research NWO. The Green Bank Observatory is a facility of the National Science Foundation operated under cooperative agreement by Associated Universities, Inc. The Arecibo Observatory is operated by SRI International under a cooperative
agreement with the NSF (AST-1100968), and in alliance with Ana~G.~M\'endez-Universidad Metropolitana
and the Universities Space Research Association.
The Parkes radio telescope is part of the Australia Telescope National Facility 
which is funded by the Australian Government for operation as a National Facility managed by CSIRO. 
This work was supported by the MPG funding for the Max-Planck Partner Group.
Part of this research was carried out at the Jet Propulsion Laboratory, California Institute of Technology, under a contract with the National Aeronautics and Space Administration (NASA).
Work at NRL is supported by NASA.
We acknowledge financial support from ``Programme National de Cosmologie and Galaxies'' (PNCG) and from ``Programme National Gravitation, R\'ef\'erences, Astronomie, M\'etrologie'' (PNGRAM) funded by CNRS/INSU-IN2P3-INP, CEA and CNES, France.
We acknowledge financial support from the project ``Opening a new era in pulsars and
compact objects science with MeerKat'' in the context of INAF grant ``SKA-CTA 2016''.
Part of this research was funded by the Australian Research Council Centre of Excellence for Gravitational Wave Discovery (OzGrav), CE170100004. 
The Flatiron Institute is supported by the Simons Foundation.
The computation was performed using the \textsc{Hercules} cluster at the Max Planck Computing and Data Facility, Garching,
 the \textsc{Dirac} cluster in KIAA and the \textsc{Tianhe II} supercomputer at Guangzhou.

RNC acknowledges the support of the International Max Planck Research School Bonn/Cologne and the Bonn-Cologne Graduate
School for part of this work.
KJL is supported by the National Natural Science Foundation of China (Grant No. 11373011).
KJL is supported by XDB23010200, National Basic Research Program of China, 973 Program, 2015CB857101 and NSFC U15311243, 11690024. We are also supported by the MPG funding for the Max-Planck Partner Group.
PL acknowledges the support of the International Max Planck Research School Bonn/Cologne. 
GD and KL acknowledge financial support by the European Research Council (ERC) for the ERC Synergy Grant BlackHoleCam under contract no. 610058.
ZA, AB, SB-S, SC, JMC, PD, TD, EF,  NG-D, PG, MTL, TJWL, ANL, DRL, RSL, DRM, MAM, STWM, CMFM, DJN, NTP, TTP, SMR, XS, JS, IS, DRS, KS, JKS, SRT, RvH are members of the  NANOGrav Physics Frontiers Center, which is supported by NSF award 1430284.
GJ and GS acknowledge support from the Netherlands Organisation for Scientific Research NWO (TOP2.614.001.602).
JW is supported by Qing Cu Hui of Chinese Academy of Sciences (CAS).
AS is supported by a University Research Fellowship of the Royal Society.
SAS acknowledges funding from the European Research Council (ERC) under the European Union's Horizon 2020 research and innovation programme (grant agreement No 694745; PI B.~W.~Stappers). 
SO was supported by the Alexander von Humboldt Foundation and acknowledges Australian Research Council grant Laureate Fellowship FL150100148.
RvH acknowledges support by NASA through Einstein Fellowship grant PF3-140116.
WWZ is supported by the Chinese Academy of Science Pioneer Hundred Talents Program and the Strategic Priority Research Program of the Chinese Academy of Sciences\ Grant No. XDB23000000. 
We thank Alessandro Ridolfi for a careful review of this paper and useful comments
as well as Agn\`es Fienga and Mickael Gastineau for useful pointers regarding
the use of \textsc{calceph}
and the anonymous referee for helpful comments and suggestions.

\bibliography{/Users/Cavas/Documents/PhD/Papers/00Refs_standard}            
\bibliographystyle{mnras}
\newpage
\parbox{.4\textwidth}{\small\emph{
$^{1}$Max-Planck-Institut f{\"u}r Radioastronomie, Auf dem H{\"u}gel 69, 53121 Bonn, Germany\\
$^{2}$Kavli institute for Astronomy and Astrophysics, Peking University, Beijing 100871,P.R.China\\
$^{3}$Jodrell Bank Centre for Astrophysics, School of Physics and Astronomy, The University of Manchester, Manchester M13 9PL,UK\\
$^{4}$Cahill Center for Astrophysics, California Institute of Technology, MC 249-17, 1200 E. California Boulevard, Pasadena, CA, 91125, USA\\
$^{5}$X-Ray Astrophysics Laboratory, NASA Goddard Space Flight Center, Code 662, Greenbelt, MD 20771, USA\\
$^{6}$Centre for Astrophysics and Supercomputing, Swinburne University of Technology, PO Box 218, Hawthorn, VIC 3122, Australia\\
$^{7}$ASTRON, the Netherlands Institute for Radio Astronomy, Postbus 2, 7990 AA, Dwingeloo, The Netherlands \\
$^{8}$International Centre for Radio Astronomy Research, Curtin University, Bentley, WA 6102, Australia\\
$^{9}$Cornell Center for Advanced Computing, Cornell University, Ithaca, NY 14853, USA\\
$^{10}$Cornell Center for Astrophysics and Planetary Science, Cornell University, Ithaca, NY 14853, USA\\
$^{11}$INAF - Osservatorio Astronomico di Cagliari, via della Scienza~5, I-09047 Selargius (CA), Italy\\
$^{12}$Department of Physics and Astronomy, West Virginia University, Morgantown, WV 26506, USA\\
$^{13}$Center for Gravitational Waves and Cosmology, West Virginia University, Chestnut Ridge Research Building, Morgantown, WV 26505\\
$^{14}$Institute for Gravitation and the Cosmos, Department of Physics, The Pennsylvania State University, University Park, PA 16802, USA\\
$^{15}$Astronomy Department, Cornell University, Ithaca, NY 14853, USA\\
$^{16}$ Laboratoire de Physique et Chimie de l'Environnement et de l'Espace LPC2E UMR7328, Universit{\'e} d'Orl{\'e}ans, CNRS, F-45071, Orl{\'e}ans, France\\
$^{17}$Station de Radioastronomie de Nan{\c c}ay, Observatoire de Paris, PSL University, CNRS, Universit{\'e} d'Orl{\'e}ans, 18330 Nan{\c c}ay, France\\
$^{18}$CSIRO Astronomy and Space Science, Australia Telescope National Facility, Box 76, Epping NSW 1710, Australia\\
$^{19}$National Radio Astronomy Observatory, PO Box O, Socorro, NM 87801, USA\\
$^{20}$Department of Physics, Hillsdale College, 33 E. College Street, Hillsdale, MI 49242, USA\\
$^{21}$Faculty of Science, University of East Anglia, Norwich Research Park, Norwich NR4 7TJ, UK\\
$^{22}$Department of Physics, McGill University, 3600 University St., Montreal, QC H3A 2T8, Canada\\
$^{23}$School of Mathematics, James Clerk Maxwell Building, Peter Guthrie Tait Road, University of Edinburgh, Edinburgh EH9 3FD, UK\\
$^{24}$Vancouver Coastal Health, Department of Nuclear Medicine, 899 W 12th Ave, Vancouver, BC V5Z 1M9, Canada\\
$^{25}$Department of Physics and Astronomy, University of British Columbia, 6224 Agricultural Road, Vancouver, BC V6T 1Z1, Canada\\
$^{26}$Department of Astrophysics/IMAPP, Radboud University, P.O. Box 9010, 6500 GL Nijmegen, The Netherlands \\
$^{27}$Space Science Division, Naval Research Laboratory, Washington, DC 20375-5352, USA\\
}}
\newpage
\parbox{.4\textwidth}{\small\emph{
$^{28}$Monash Centre for Astrophysics (MoCA), School of Physics and Astronomy, Monash University, VIC 3800, Australia\\
$^{29}$Jet Propulsion Laboratory, California Institute of Technology, 4800 Oak Grove Dr, M/S 67-201, Pasadena, CA 91109, USA\\
$^{30}$Department of Physics and Astronomy, Haverford College, 370 Lancaster Ave, Haverford, PA 19041\\ 
$^{31}$National Radio Astronomy Observatory, 520 Edgemont Rd, Charlottesville, VA 22903, USA\\
$^{32}$Center for Computational Astrophysics, Flatiron Institute, 162 Fifth Avenue, New York, NY 10010, USA\\
$^{33}$Physics Department, Lafayette College, Easton, PA 18042, USA\\
$^{34}$Fakult\"at f\"ur Physik, Universit\"at Bielefeld, Postfach 100131, 33501 Bielefeld, Germany\\
$^{35}$Physics Department, Texas Tech University, Box 41051, Lubbock, TX 79409, USA\\
$^{36}$Hungarian Academy of Sciences MTA-ELTE ``Extragalactic Astrophysics'' Research Group, Institute of Physics, E\"otv\"os Lor{\'a}nd University, P{\'a}zm{\'a}ny P.~s. 1/A, Budapest 1117, Hungary\\ 
$^{37}$Anton Pannekoek Institute for Astronomy, University of Amsterdam, Science Park 904, 1098 XH Amsterdam, The Netherlands\\
$^{38}$School of Physics and Astronomy, The University of Birmingham, Edgbaston, Birmingham, B15 2TT, UK\\
$^{39}$The Australian Research Council Centre of Excellence for Gravitational Wave Discovery (OzGrav)\\
$^{40}$Center for Gravitation, Cosmology and Astrophysics, Department of Physics, University of Wisconsin-Milwaukee, PO Box 413, Milwaukee, WI 53201, USA\\
$^{41}$Physics and Astronomy Department, Oberlin College, Oberlin, OH 44074, USA\\
$^{42}$TAPIR Group, MC 350-17, California Institute of Technology, Pasadena, California 91125, USA \\
$^{43}$Laboratoire Univers et Th\'eories (LUTh), Observatoire de Paris, PSL University, CNRS, Universit\'e Paris-Diderot, 92190 Meudon, France\\
$^{44}$Institute for Radio Astronomy \& Space Research, Auckland University of Technology, Private Bag 92006, Auckland 1142, New Zealand\\
$^{45}$Xinjiang Astronomical Observatory, Chinese Academy of Sciences, 150 Science 1-Street, Urumqi, Xinjiang 830011, China\\
$^{46}$National Astronomical Observatories, Chinese Academy of Science, 20A Datun Road, Chaoyang District, Beijing 100012, China\\
}}

\end{document}